\documentclass[12pt]{iopart}
\usepackage{epsfig,iopams,amssymb,mathrsfs,amsthm,graphicx,float,color}
\usepackage[active]{srcltx}
\usepackage{subfigure}
\newcommand{\N}{\mathbb{N}}

\newcommand{\set}[1]{\mathsf{#1}}
\newcommand{\grp}[1]{\mathsf{#1}}
\newcommand{\spc}[1]{\mathcal{#1}}

\def\d{{\rm d}}

\def\>{\rangle}
\def\<{\langle}
\def\kk{\>\!\>}
\def\bb{\<\!\<}
\newcommand{\st}[1]{\mathbf{#1}}

\newcommand{\map}[1]{\mathcal{#1}}

\newtheorem{theo}{Theorem}

\newtheorem{lemma}{Lemma}
\newtheorem{prop}{Proposition}

\newtheorem{eg}{Example}

\def\Proof{{\bf Proof.~}}
\def\qed{$\blacksquare$ \newline}

\begin{document}
\title{Units of rotational information}

\author{Yuxiang Yang} 
\address{Department of Computer Science, The University of Hong Kong, Pokfulam Road, Hong Kong}
\address{HKU Shenzhen Institute of Research and Innovation,  Kejizhong 2$^{\rm nd}$ Road,    Shenzhen,  China}

\author{Giulio Chiribella\footnote{email: giulio@cs.hku.hk}} 
\address{Department of Computer Science, University of Oxford, Parks Road, Oxford, UK}
\address{Canadian Institute for Advanced Research, CIFAR Program in Quantum Information Science, Toronto, ON M5G 1Z8}

\author{Qinheping Hu} 
\address{Department of Computer Science, University of Wisconsin-Madison, 1210 W. Dayton St.
Madison, WI 53706-1613. }

\begin{abstract}
Entanglement in   angular momentum degrees of freedom is a precious  resource for quantum metrology and control.  
  Here we study the conversions of this resource,  focusing on Bell pairs of spin-$J$ particles,  
   where one particle  is used to probe unknown  rotations and the other particle is used as  reference.   
     When a large number of pairs are given, we show that  every rotated spin-$J$ Bell state can be reversibly converted into an equivalent number of rotated spin one-half Bell states, at a rate  determined by  the quantum Fisher information. 
        This result provides the foundation for the definition of an elementary unit of information about rotations in space,  which we call the \emph{Cartesian refbit}.  
     In the finite copy scenario,   we design machines that approximately  break down   Bell states  of higher spins into  Cartesian refbits,  as well as machines  that approximately  implement   the inverse process.   In addition, we  establish a quantitative link between the conversion of Bell states and  the simulation of unitary gates, showing that the fidelity of probabilistic state conversion provides upper and lower bounds on the fidelity of deterministic gate simulation.   The result holds not only for rotation gates, but also to all sets of gates that form finite-dimensional representations of compact groups.      For rotation gates,  we show  how  rotations on a system of given spin can simulate rotations  on a system of different spin.    
\end{abstract}

\maketitle

\section{Introduction}
Quantum states that  encode information  in  the angular momentum degree of freedom are a valuable resource for quantum metrology \cite{giovannetti-lloyd-2004-science,toth-2012-pra} and communication  \cite{dambrosio-2012-natcomm,vallone-2014-prl}. 
      But depending on the task at hand, certain states  can be more useful than others. 
    In situations where quantum communication is a scarce resource, it is natural to prefer entangled states that convey precise information  with the smallest number of particles.  In situations where joint operations are challenging to implement, it is more preferable to encode information into product states, even if such encoding requires an overhead in the number of particles.   When different tasks are composed, it becomes useful to switch from one encoding to another: for example, one may want to first transfer directional information from a sender  to a receiver  (using the minimum amount of quantum communication) and then to broadcast the information from the receiver to a number of local users (using an encoding that allows to read out the information locally).         A device that implements the conversion between one encoding and the other  acts as  an ``adapter", which converts information from a form that is easier to transmit to a form that is easier to read out. 
    
    In this  paper    we  focus on   the conversions of maximally entangled  bipartite states, also known as Bell states.  Bell states of systems with definite angular momentum are faithful carriers of information about rotations in space: when a rotation $R$ is applied locally on one part of a Bell state $|\Phi\>$, the resulting  Bell state $|\Phi_R\>=  (R\otimes I)  |\Phi\>$  is in one-to-one correspondence with  $R$.   Even more specifically,  the Bell states are  optimal for  probing  rotations  among the states of systems with definite value of the angular momentum   \cite{acin-jane-2001-pra,chiribella-dariano-2005-pra}. Bell states are also optimal for the task of storing/retrieving rotation gates   \cite{bisio-chiribella-2010-pra} and for correcting errors due to the lack of a shared reference frame of Cartesian directions \cite{bartlett-rudolph-2009-njp}.   
    
     The conversion of Bell states is the  paradigmatic example of optimal conversions of quantum reference frames  \cite{bartlett-rudolph-2007-rmp}. In this paper we  study the Bell state example in depth, determining  the convertibility conditions  and highlighting their physical meaning.  This work illustrates and complements the general theory of asymmetry as a resource \cite{vaccaro-anselmi-2008-pra,gour-spekkens-2008-njp,gour-marvian-2009-pra,skotiniotis-gour-2012-njp,marvian-spekkens-2013-njp,marvian-spekkens-2014-pra-modes,marvian-spekkens-2014-pra-asymmetry}, offering a concrete case study that can be used  for further generalizations.  Specifically, we investigate the  problem of converting $N$ copies of a spin-$J$ Bell state into $M$ copies of a spin-$K$ Bell state, while preserving the information about  local rotations.       One example of this type of conversions  is  the cloning of Bell states  \cite{lamoureux-navez-2004-pra,karpov-navez-2005-pra,chiribella-yang-2014-njp},  corresponding to the case  $J  = K$ and $M>N$.   
In the large $N$ limit,  we show that  a deterministic and reversible conversion can be achieved whenever the quantum Fisher information is conserved at the leading order.     
Our result supports a conjecture by Marvian and Spekkens \cite{marvian-spekkens-2014-pra-asymmetry}, who suggested that, under the validity  of certain symmetry conditions, the  conservation of the quantum Fisher information  should  be sufficient for an asymptotically reversible conversion of quantum reference frames.   

Since all the  Bell pairs of spin-$J$ particles  are asymptotically interconvertible with each other, we can regard the spin-1/2 Bell pair as the standard unit of information about rotations in space, or equivalently, about Cartesian reference frames.  Borrowing a term introduced by van Enk \cite{vanenk-2005-pra,vanenk-2006-pra}, we call  the spin-1/2 Bell pair a \emph{Cartesian refbit}---a bit of Cartesian reference frame.   

We then consider two categories of machines:   One category  of machines break down Bell states into Cartesian refbits. We name such machines \emph{quantum analysers}.  The other category of machines perform the opposite conversion, merging groups of Cartesian refbits into  Bell states of higher angular momenta. We name these machines    \emph{quantum synthesisers}.  

Decomposing/recomposing quantum states into/from basic units of reference frame has a number of interesting applications.   For example,    quantum  analysers can be used to distribute  directional information to multiple receivers:  Using a quantum analyser, a high-precision gyroscope can be broken down into  a number of elementary  gyroscopes, each carrying a unit of directional information. In this way, the original information can be distributed to multiple receivers, who can then perform local measurements.  Essentially, the quantum analyser takes care of the hard part in the readout and redistributes the information in a form that can be accessed locally.    Quantum synthesizers, instead, can be used to compress directional information into a more compact form that is useful for storage into the quantum memory of a quantum computer or for  transmission via a quantum communication line.

In the non-asymptotic scenario we find  that quantum analysers exhibit a number of peculiar properties. For example,   we find that  individual    Bell states are   ``unbreakable", meaning that no quantum analyser can convert a \emph{single} Bell state into Cartesian refbits  with  high level of accuracy.    This fact is in stark contrast with the situation for spin-$J$ coherent states \cite{radcliffe-1971-jpa,arecchi-1972-pra}, which  can be reversibly  broken down into   $M= 2J$ spin-$1/2$ coherent states.   The contrast is worth highlighting  because, among the states of systems with definite angular momentum,  the spin coherent states are the best carriers of  information  about individual directions  \cite{book-holevo-probabilistic}, while the Bell states are the best carriers of information about Cartesian reference frames  \cite{acin-jane-2001-pra,chiribella-dariano-2005-pra}.    The contrast between spin coherent and Bell states highlights a fundamental difference between the communication of a single direction  and the communication of a full Cartesian frame: while the best states  for communicating individual directions can be broken down into elementary units, the best states for communicating Cartesian frames cannot. Heuristically, the difference arises  from the particular way in which Cartesian frames are encoded into  Bell states: rather than localizing the information about three directions onto three different systems,  the Bell state concentrates the information  into one entangled pair.     Such a      way to pack  information is system-specific, and  systems with different spin correspond to different, inequivalent encodings. 

Besides the conversions of Cartesian reference frames, our results determine  how rotation gates on a system of a given spin can be simulated by rotation gates on a system with  different spin.  For example,  imagine the scenario where   a black box  performs  an unknown  rotation on a spin-1/2 particle.  By using the blackbox for $N$ times, a machine can simulate  the rotation of a  higher angular momentum.  But how large should $N$ be in order to reproduce the desired rotation with high accuracy?  And how many times can the machine execute the rotation?  
  To address  these questions we derive a general result,  bounding the average performance of    \emph{deterministic} gate conversion  with the performance of a \emph{probabilistic} Bell state conversion.  
Specifically, we show that the two fidelities satisfy the relation    
  \begin{equation}
 \left(  F^{\rm prob}_{\rm Bell} \right)^2  \le   F^{\rm det}_{\rm gate}\le  F^{\rm prob}_{\rm Bell}    \, ,
  \end{equation} 
  valid not only for rotations but also for every compact group of unitary transformations. 
     As a consequence, we show that a gate simulation can be achieved deterministically with high fidelity if and only if  the corresponding state conversion can be achieved probabilistically with high fidelity.  Once this fact is established,  every result  on the probabilistic conversion of Bell states can be  translated into a result  on the deterministic simulation of rotation gates.  For example, we find that  a single rotation of a spin-$J$ system cannot be used to simulate rotations on spin-1/2 systems. 
     Our results provide tools that can be applied also beyond the problem of simulating  rotation gates. In a broad perspective, they contribute to the study of quantum machines capable to  automatically learn how to perform desired tasks, such as learning an unknown unitary gate \cite{bisio-chiribella-2010-pra} or learning a quantum measurement \cite{bisio-dariano-2011-pla}. 

The paper is organised as follows. In Section \ref{sec:general}, we introduce  the general framework. 
   The deterministic conversions of Bell states are studied in Section \ref{sec:deterministic}, where we introduce the notion of Cartesian refbit.      The probabilistic conversions are then  studied in Section \ref{sec:probabilistic}.  Then, we move to the problem of analyzing/synthesizing Bell states into/from Cartesian refbits.       In Section \ref{sec:fission}, we focus on the task of breaking angular momentum  Bell states into Cartesian refbits. In Section \ref{sec:fusion}, we focus on the dual task of merging Cartesian refbits into   Bell states of higher angular momenta.    Section \ref{sec:gate} addresses the simulation of rotation gates and its relation to spin  conversions, providing general results valid for arbitrary groups of unitary gates. Finally, the conclusions are drawn  in section \ref{sec:discussion}.  The technical proofs are provided in the appendices, which can be skipped at a first reading.

\section{Bell state conversions}\label{sec:general}
In this section we introduce the general  problem of converting  angular momentum Bell states, defining the notation and the relevant figures of merit used in the paper.

\subsection{The task}
Imagine that an experimenter has access to a black box performing an unknown rotation on a quantum system with definite  angular momentum, specified by the quantum number $J$.   Let us denote  by   $g  \in  \grp {SO}(3)$ the rotation, and by $U_{g,J}$ is the  unitary matrix that represents  the rotation on the system's Hilbert space. 

\begin{figure}  
\begin{center}
  \includegraphics[width=0.6\linewidth]{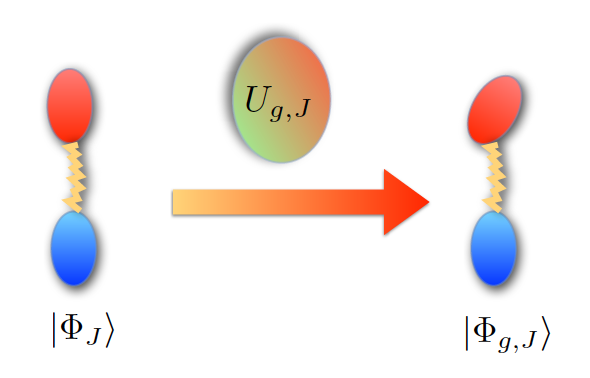}
  \end{center}
\caption{\label{fig:field}
  {\bf Encoding a local rotation into a  Bell state.}  A spin $J$ system  (in red), acting as  a probe, is entangled with another spin $J$ system (in blue), acting as a reference.   The two systems are initially in the standard Bell state  $|\Phi_J\>$. Then, the probe undergoes the unknown rotation   $g$ and the state of the composite system is transformed into a rotated Bell state  $|\Phi_{g,J}\>$.}
\end{figure}

For many applications, it is useful to imprint the rotation into the state of a quantum system.  For example, the application could be to communicate the direction of three  Cartesian axes  \cite{chiribella-dariano-2004-prl,bagan-baig-2004-pra,hayashi-2006-pla}, to sense an unknown magnetic field  \cite{toth-2012-pra}, or to  store the rotation in the the memory of a quantum computer \cite{bisio-chiribella-2010-pra},  to correct for an error   \cite{bartlett-rudolph-2009-njp,bisio-chiribella-2010-pra}, or to generate a quantum program for a programmable measurement device \cite{marvian-2012-arxiv,ahmadi-jennings-rudolph-2013-njp}. For a single  use of the black box, the optimal way to imprint the rotation gate     is illustrated in  Figure  \ref{fig:field}.  Explicitly, one has to  
\begin{enumerate}
\item prepare a pair of spin-$J$ systems in the standard Bell state 
\begin{eqnarray}  
  |\Phi_J\>    :=     \frac{ \sum_{m=-J}^J  \, |J,m\>\otimes |J,m\>}{\sqrt{2J+1}} \, 
  \end{eqnarray} 
 where    $\{  |J,m\>~|~ m=  -J, \dots,  +J\}$ are  the eigenvectors of the $z$ component of the angular momentum operator.
 \item  let the  first system  undergo the rotation, so that the standard Bell state is transformed    into the rotated Bell state  
\begin{eqnarray}\label{state def}
|\Phi_{g,J}\>:=      \left(    U_{g,J}  \otimes I_J \right)  \,    |\Phi_J\> \, ,   
\end{eqnarray}
where $I_J$ is the identity matrix. 
\end{enumerate}  
  
  By repeating this procedure on $N$ pairs, the experimenter can generate $N$ identical copies of the rotated Bell state $|\Phi_{g,J}\>$.     At this point, the $N$ copies   represent a physical token of the information about the  rotation. Bell states corresponding to the same rotation  but to different values of the angular momentum represent different types of tokens.

In the following we consider the  task  of converting  one type of token into another.  
  Precisely, we will search for the optimal process that   transforms $N$ copies of a rotated spin-$J$ Bell state   into $M$ approximate copies of a  rotated spin-$K$ Bell state, while preserving the information about the rotation. Ideally, we aim at implementing the transformation
\begin{eqnarray}\label{ideal}
|\Phi_{g,J}\>^{\otimes N}\to|\Phi_{g,K}\>^{\otimes M}  \,  ,  \qquad \forall g\in\grp {SO} (3) \, .
\end{eqnarray}
In most cases, such transformation cannot be implemented perfectly.    We will refer to the task of approximating the desired  transformation  as \emph{``converting $N$ copies of a rotated spin-$J$ Bell state into $M$ copies the corresponding spin-$K$ Bell state"}. Implicitly, it is understood  that  the rotation  $g$ in the input Bell state $|\Phi_{g,J}\>^{\otimes N}$ is unknown and therefore the conversion mechanism should be independent of $g$.

\subsection{Optimal quantum machines} 

In this paper we consider two ways of converting Bell states: by  deterministic operations and by probabilistic operations.   
A deterministic machine   is described by a quantum channel  (completely positive trace-preserving map) $\map{C}$, transforming the state of the $N$ input pairs into the state of the $M$ output pairs. The  machine converts the $N$-copy   input   state  $|\Phi_{g,J}\>^{\otimes N}$ into the (generally mixed) output state $\map{C}\left(|\Phi_{g,J}\>\<\Phi_{g,J}|^{\otimes N}\right)$.  Note that, in general we allow the machine to perform global operations jointly on all the input systems.   The performance of the  machine is  measured  by the average fidelity between the output state and the desired $M$-copy state, namely
\begin{eqnarray}\label{Fdef}
\fl \qquad  F^{\rm det}_{\rm Bell}  \Big[  |\Phi_{g,J}\>^{\otimes N}\to |\Phi_{g,K}\>^{\otimes M}  \Big]=\int \d g ~\<\Phi_{g,K}|^{\otimes M}\map{C}\left(|\Phi_{g,J}\>\<\Phi_{g,J}|^{\otimes N}\right)|\Phi_{g,K}\>^{\otimes M}.
\end{eqnarray}

A probabilistic machine  is described by a quantum operation (completely positive trace-non-increasing map) $\map{M}$. The occurrence of  the probabilistic transformation $\map{M}$ is heralded by the outcome of a quantum  measurement. We call this outcome the ``successful outcome", meaning that, when the outcome occurs,  the machine  produces an output according to the intended map $\map M$. 
In such a case, the output state is
\begin{eqnarray}
\rho_g'   =   \frac{\map{M}\left(|\Phi_{g,J}\>\<\Phi_{g,J}|^{\otimes N}\right)}{\Tr \left[  \map{M}\left(|\Phi_{g,J}\>\<\Phi_{g,J}|^{\otimes N}\right) \right]}  
\end{eqnarray}
and the probability of success is  
\begin{eqnarray}
p({\rm succ}|g)   =  \Tr \left[ \map{M}\left(|\Phi_{g,J}\>\<\Phi_{g,J}|^{\otimes N}\right) \right]   \, .
\end{eqnarray}
Conditionally on the occurrence of the successful outcome, the performance of the probabilistic machine is  evaluated by the  average fidelity between the output state and the desired $M$-copy state,     namely 
\begin{eqnarray}
F^{\rm prob}_{\rm Bell} \Big [  |\Phi_{g,J}\>^{\otimes N}\to |\Phi_{g,K}\>^{\otimes M}\Big]=\int p\left(\d g|{\rm succ}\right)~  \<\Phi_{g,K}|^{\otimes M}  
\,  \rho_g'  \, |\Phi_{g,K}\>^{\otimes M}
 \, ,   \qquad \qquad
\end{eqnarray}
where  $ p\left(\d g|{\rm succ}\right)$ is the conditional probability distribution for the rotation $g$.  Specifically, the probability distribution can be expressed as  $p(\d g|{\rm succ})=p({\rm succ}|g)\,\d g/p_{\rm succ}$, where  $\d \, g $ is the normalized Haar measure and  
\begin{eqnarray}\label{p-succ}
p_{\rm succ}=\int \d g~\Tr\left[\map{M}\left(|\Phi_{g,J}\>\<\Phi_{g,J}|^{\otimes N}\right)\right]
\end{eqnarray}
is the total success probability.
Combining the above relations, the probabilistic fidelity  reduces to
\begin{eqnarray}\label{Fdef-prob}
\fl  \qquad F^{\rm prob}_{\rm Bell}  \Big[     |\Phi_{g,J}\>^{\otimes N}\to |\Phi_{g,K}\>^{\otimes M}   \Big]&=\frac{\int \d g ~ \<\Phi_{g,K}|^{\otimes M}\map{M}\left(|\Phi_{g,J}\>\<\Phi_{g,J}|^{\otimes N}\right)|\Phi_{g,K}\>^{\otimes M}}{\int \d g~\Tr\left[\map{M}\left(|\Phi_{g,J}\>\<\Phi_{g,J}|^{\otimes N}\right)\right]} \, . 
\end{eqnarray}
This expression will be often used in our analysis.  In the end of the paper we will  show that  the probabilistic fidelity in Eq. (\ref{Fdef-prob}) provides bounds on a gate simulation task, where the goal is to simulate $M$ uses of a rotation on spin-$K$ system with $N$ uses of the same rotation on a spin-$J$ system.

\section{Deterministic conversions}\label{sec:deterministic}
In this section we characterize the conversions of Bell states that can be achieved deterministically. We first consider the simplest instance of the problem,  involving a single input Bell state and a single output Bell state.  Then, we move to conversions  involving asymptotically many copies.   In the asymptotic setting, we  identify  the conservation of the quantum Fisher information as the necessary and sufficient condition for a faithful conversion. 
\subsection{Single-copy conversions}\label{subsec:singlecopy}  

Let us start from the simple case where the input of the conversion  is a single copy of  a spin-$J$ Bell state. In this case, the symmetry of the problem   allows us to identify the optimal conversion process and to give an  analytical expression for the fidelity.   
Here we focus on the results and on their physical interpretation, while the technical details  are provided in \ref{app:singlecopy}.

In the single-copy case, it turns out that  deterministic and probabilistic operations perform equally well, no matter how small is the probability of success.    The optimal fidelity for converting one spin-$J$ Bell state into one spin-$K$ Bell state is 
   \begin{eqnarray}\label{JFK} 
F_{\rm Bell}  \Big [   |\Phi_{g,J}\>\to |\Phi_{g,K}\> \Big]  =     \frac{2J+1}{(2K+1) \, (2|J-K|+1)} \, .
\end{eqnarray}
An important observation is that the conversion is never perfect, except in the trivial cases $J=K$ and/or $K=0$.    
 For all the other values of $J$ and $K$ the fidelity satisfies the bound 
 \begin{eqnarray}F_{\rm Bell}  \Big[   |\Phi_{g,J}\> \to |\Phi_{g,K}\>  \Big]  \le  75\% \,, 
 \end{eqnarray}   
 where the equality is attained by setting $J= 1$ and $K  = 1/2$.    For large $J$ or large  $K$,  the fidelity tends to zero  as the difference $|J-K|$ becomes large. 
 
   
   \begin{figure}[t!]
\centering
      \includegraphics[width=0.6\textwidth]{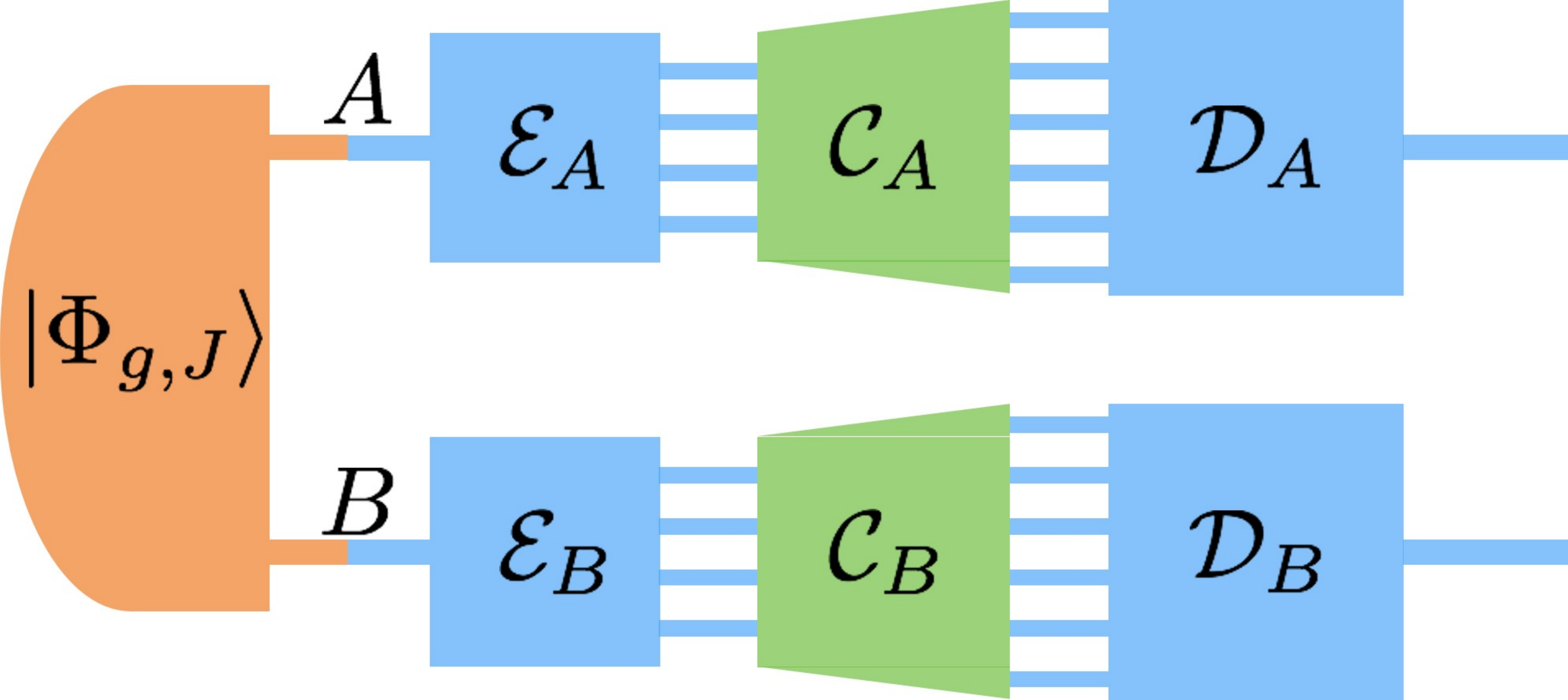}\caption{{\bf   Optimal single-copy Bell state converter.}   The figure illustrates the action of  an optimal machine converting a spin-$J$ Bell state $|\Phi_{g,J}\>$ into a spin-$K$ Bell state $|\Phi_{g,K}\>$ for the $J\le K$ case. Two identical sequences of operations are applied to each of the two  subsystems constituting the Bell pair:  first, the encoding channel $\map{E}$ embeds the spin-$J$ system into a system of $2J$ spin-1/2 systems (qubits).     Then,    the universal cloning machine  $\map{C}$  optimally turns $2J$ qubits into $2K$ qubits.  Finally, the decoding channel  $\map{D}$  merges $2K$ qubits into a single spin-$K$ system. A similar sequence of operations allows us to achieve conversions with  $J>K$, the only difference being that on has to replace  the universal cloning  with a universal discarding, corresponding to the partial trace over $2J-2K$ qubits.  In both cases, the optimal conversion only requires local operations on the two subsystems of the Bell pair.}
       \label{fig:clontrace}
\end{figure}

 The optimal conversion process has an intuitive physical realization. The idea is that a single  spin-$J$ system can be faithfully  encoded into  a  system of $2J$  spin-1/2 particles, whose state is constrained to be in the symmetric subspace \cite{harrow2013church}.      
When  $J$ is smaller than $K$, the initial $2J$ particles can be converted into $2K$ particles by  using the universal quantum cloning machine \cite{werner-1998-pra}.     When  $J$ is larger than $K$, one has to discard $2(J-K)$ of the particles.  In both cases, the protocol produces  $2K$ spin-$1/2$ particles in the symmetric subspace.    Thanks to this fact,  the  $2K$ particles  can be   transformed into a single spin-$K$ system by a suitable decoding operation. The overall protocol is illustrated in Figure \ref{fig:clontrace}.

Figure \ref{fig:clontrace} shows that the   optimal Bell state conversion is achieved by local operations, performed independently on the two input spins. For $K>J$, there is an interesting connection with the cloning problem considered in Refs. \cite{cwiklinski2012region,studzinski2014group}, where the aim is to locally clone the correlations between a system and a reference. It turns out that the universal cloning machine is optimal both for the local cloning problem of Refs. \cite{cwiklinski2012region,studzinski2014group} and for the problem of converting Bell states.  In a sense, the conversion of the Bell state $ |\Phi_{g,J}\> $ into the Bell state $|\Phi_{g,K}\>$ can be viewed as the local cloning of correlations, with cloning operations performed both on the system and on the reference.  

One may wonder whether this is a generic feature of Bell state conversions. We can imagine that, for every Bell pair, one spin is in Alice's laboratory and the other is in Bob's laboratory.  Then the question is:  can Alice and Bob achieve the optimal Bell state conversion by performing local operations in their laboratories and, possibly, coordinating their operations through the communication of classical messages?    Interestingly, this is not the case for $N >1$ or $M>1$: later in the paper we will see that, in general,  joint operations  are necessary in the multicopy scenario.   



\subsection{Asymptotic  conversions}\label{subsec:asymptotic}

Here we consider asymptotic conversions  where one is given a large number of spin-$J$ pairs, each pair in the same rotated  Bell state. The goal is   to produce as many many spin-$K$ pairs as possible, under the condition that the joint state of all pairs should resemble $M$ perfect copies of the rotated spin-$K$ Bell pair, with an  error vanishing in the asymptotic limit. 
     In the asymptotic scenario, it turns out that  Bell states with different angular momenta can be interconverted reversibly, as shown by the following Theorem. 
     \begin{theo}[Deterministic  Bell state conversion]\label{thm:generaldet}
If the condition
\begin{eqnarray}\label{fishercond} 
\Big  |  MK(K+1)   -   NJ(J+1)   \Big| =  \Delta   
\end{eqnarray}
holds with    $\Delta  =   O(  N^{1-\alpha})$ for some $\alpha >0$,  then there exists a deterministic machine that reversibly transforms  $N$ copies of the spin-$J$ Bell state $|\Phi_{g,J}\>$ into $M$ copies of the spin-$K$ Bell state $|\Phi_{g,K}\>$ with   error  vanishing as $[\Delta/N J(J+1)]^2$ in the large $N$ limit. 
\end{theo}
The idea of the proof is to decompose the $N$-copy input states into a superposition of eigenstates with definite values of the quantum number of  the total angular momentum.   When this is done, it turns out that the quantum number of the total angular momentum is asymptotically distributed as a Gaussian  with variance $NJ (J+1)/3$, times a polynomial prefactor.   Specifically, for integer $NJ$ one has  the decomposition  
\begin{eqnarray}\label{inputdecompj}
 |\Phi_{g,J}\>^{\otimes N}=\bigoplus_{j =0}^{NJ} \sqrt{p^{(N,J)}_{j}} \left|\Psi_{g,j}^{(N,J)}\right\> \, , 
\end{eqnarray}
where  $j$ is the quantum number of the total angular momentum,  $|\Psi_{g,j}^{(N,J)} \>$  is an  eigenstate of  the square of the total angular momentum operator,  and $p_j^{(N,J)}$  is a probability distribution, asymptotically equal to
\begin{eqnarray}\label{papprox}
\fl \qquad p^{(N,J)}_{j}=\sqrt{\frac{27(2j+1)^{4}}{8\pi N^{3}J^{3}(J+1)^{3}}}\exp\left[-\frac{3j^{2}}{2NJ(J+1)}\right]\left[1-O\left(\frac{1}{N (J+1)}\right)\right]
\end{eqnarray}
(see \ref{app:inputdecomp} and \ref{app:asymptotic} for the derivation of Eqs.  (\ref{inputdecompj}) and (\ref{papprox}), respectively).

    The same decomposition holds for the $M$-copy output space, except that the variance of the Gaussian is $MK(K+1)/3$, instead of $NJ(J+1)/3$.    To convert the input state into the output state, we use  a transformation that preserves the total angular momentum, while transforming the state $  |\Psi_{g,j}^{(N,J)}\>$ into the state $|\Psi_{g,j}^{(M,K)}\>$ for every value of $j$.    The conversion has high-fidelity if the Gaussian distributions of the input and output states are close, which happens when $MK(K+1)$ is equal to  $NJ(J+1)$ at the leading order. The proof details can be found in  \ref{app:asymptodet}.   

 Theorem \ref{thm:generaldet} tells us that spin-$J$ Bell states can be reversibly converted into spin-$K$ Bell states, provided that  the two quantities $MK(K+1)$ and $N J (J+1)$ are close to each other.     In particular, this means that the ratio between the number of output and input copies  grows asymptotically as 
\begin{eqnarray}\label{fisher}
\frac MN    =  \frac{J  (J+1)}{K(K+1)}   +\epsilon  \, ,\end{eqnarray}
where $\epsilon$ vanishes as $\Delta/N$.    

Note that, in general,  the conversion of rotated Bell states cannot be achieved by local operations. For local operations, the theory of pure state entanglement \cite{bennett1996concentrating}  implies that the ratio $M/N$ must be smaller than or equal to $\log (2J+1)/\log(2K+1)$.  This means that conversion of rotated Bell states requires global operations whenever  $\log (2J+1)/\log(2K+1)$  is smaller  than $J(J+1)/[K(K+1)]$.    When this is the case, the conversion of rotated Bell states requires global operations, capable to generate entanglement, while preserving the information about the rotation.

\subsection{Conservation of the Fisher information} 

The condition (\ref{fisher})  has an intuitive interpretation in terms of the amount of  information carried by the input and output states.       Suppose that  rotation $g$ is parametrized  in terms of three rotation angles, corresponding to rotations around the axes $x$, $y$, and $z$.   To discover the three rotation angles  $\boldsymbol \theta =     (\theta_x,  \theta_y,\theta_z)$, it is convenient to  use an unbiased measurement, that is, a measurement that on average returns the correct angles.  The precision of the measurement can be quantified by  the covariance matrix $C_{\boldsymbol \theta}$, defined as  
 \begin{eqnarray}
 \big [C_{\boldsymbol \theta} \big]_{ij}  :  =  \int   \d  \boldsymbol \theta  \,  p(  \hat {\boldsymbol \theta}   |  \boldsymbol  \theta)   \,  (\hat \theta_i  -  \theta_i)  \,  (\hat \theta_j   - \theta_j) \, ,   
 \end{eqnarray} 
 where   $\hat {\boldsymbol \theta} =     (\hat \theta_x,  \hat \theta_y,  \hat  \theta_z)$  are the measured angles and $p(  \hat {\boldsymbol \theta}   |  \boldsymbol  \theta) $ is the conditional probability distribution of measuring $\hat {\boldsymbol \theta} $ when the true angles are $ {\boldsymbol \theta}$. The covariance matrix can be bounded in terms of the quantum Fisher information matrix  $F_{\boldsymbol \theta}$, which for a pure state $|\Psi_{\boldsymbol \theta} \>$ is defined as  
  \begin{eqnarray}
 \big [{\rm QFI}_{\boldsymbol \theta} \big]_{ij}  :  =  4   {\rm Re}  \Big [    \<   \Phi_{\boldsymbol \theta  ,  i}    |   \Phi_{\boldsymbol \theta  , j} \>   -   \<   \Phi_{\boldsymbol \theta  }    |   \Phi_{\boldsymbol \theta  , i}  \>  \<   \Phi_{\boldsymbol \theta  ,  j}    |   \Phi_{\boldsymbol \theta } \> \Big]    \, ,
 \end{eqnarray} 
  where we used the notation   $ |   \Phi_{\boldsymbol \theta  , i} \>  :=   \frac{\partial}{\partial  \theta_i}  |\Psi_{\boldsymbol \theta}\> $.  
   The bound on the covariance matrix, known as  the quantum Cram\'er-Rao bound \cite{book-helstrom-1976-quantum,book-holevo-probabilistic,braunstein-caves-1994-prl},  has  the form 
 \begin{eqnarray}
 C_{\boldsymbol \theta}  \ge  {\rm QFI}_{\boldsymbol \theta}^{-1}  \, ,  
 \end{eqnarray}   
 where ${\rm QFI}_{\boldsymbol \theta}^{-1} $ denotes the inverse of the matrix ${\rm QFI}_{\boldsymbol \theta}$, and the notation $A  \ge B$ means that the all the eigenvalues of the matrix $A-B$ are positive or zero.   In particular, the quantum Cram\'er-Rao bound implies that the variance for the measurement of the angles $\theta_x$  $\theta_y$, and $\theta_z$ are lower bounded by the diagonal entries of the inverse quantum Fisher information matrix.  
 
For the spin-$J$ Bell states the quantum Fisher information matrix is independent of $\boldsymbol \theta$ and is given by  \cite{chiribella-chao-2014-arxiv}

\begin{eqnarray}
 {\rm QFI}    =   \frac{ 4NJ(J+1) } {3}    \left(\begin{array}{ccc} 1  &  0  &  0  \\    0  &  1  &  0  \\  0&  0  &  1\end{array}\right)   \, .  
\end{eqnarray}
Since the quantum Fisher information matrix is proportional to the identity, we can simply focus on the proportionality constant $  4NJ(J+1)/3$ and refer to it as the ``quantum Fisher information''.     We can now give an intuitive interpretation to the   condition (\ref{fisher})  on the asymptotic convertibility of Bell states. The condition is the    \emph{(approximate)  conservation of the Fisher information} from the input to the output:  
   if  the  quantum Fisher information of the input is approximately equal to the quantum Fisher information of the output, then  the transition is asymptotically possible and can be implemented reversibly.

  \subsection{The Marvian-Spekkens conjecture}
For families of pure states generated by  rotations, the conservation of the quantum Fisher information is equivalent to the conservation  the covariance matrix of the angular momentum operator.  This condition was  identified  by  Marvian and Spekkens \cite{marvian-spekkens-2014-pra-asymmetry}   as  a necessary requirement  for  the reversible, asymptotic convertibility of  pure states.  In the same work, Marvian and Spekkens  conjectured that the conservation of the covariance matrix should also be  sufficient, provided that two additional symmetry requirements are satisfied.   
 In our settings, these requirements are trivial and therefore the Marvian-Spekkens conjecture becomes that the conservation of the Fisher information is necessary and sufficient for an asymptotically reversible conversion. Theorem \ref{thm:generaldet} proves the validity of this conjecture in the case of rotated Bell states.

In the Bell state case, we can also provide a strong converse to the Marvian-Spekkens conjecture, showing that the quality of the conversion vanishes whenever the conversion rate exceeds the value determined by the conservation of the Fisher information.   Specifically,  we prove that every deterministic machine has to satisfy the upper bound 
\begin{eqnarray}\label{asymptoticupper}
F^{\rm det}_{\rm Bell}   \Big [   |\Phi_{g,J}\>^{\otimes N}\to |\Phi_{g,K}\>^{\otimes M} \Big]\le\left[\frac{NJ(J+1)}{MK(K+1)}\right]^{\frac32}+O\left(\sqrt{\frac{N}{M^3}}\right) \,.
\end{eqnarray}
 valid for large $N$ and $M$.     The derivation of the bound is provided in  \ref{app:det-bound}.  
  According to the bound (\ref{asymptoticupper}), a deterministic machine that over-produces Bell states will incur in an error, proportional to the extent to which the conservation of the quantum Fisher information has been violated.  
   For example, a machine that produces Bell states at a quadratic rate $M    \propto N^2$ will   necessarily have  vanishing fidelity in the asymptotic limit.  

\subsection{The Cartesian refbit}

The asymptotic  convertibility of Bell states provides the foundation for the definition of an elementary unit of information about rotations in space.     As a standard unit of information, we choose the 
spin-1/2 Bell   state    $|\Phi_{g, 1/2}\>$.
There are two reasons for this choice:  
\begin{enumerate}
\item  the spin-1/2 Bell state is the \emph{best} state that carries faithful information about rotations on the {\em smallest} quantum system
\item in the asymptotic setting every spin-$J$ Bell state can be reversibly  converted into spin-1/2 Bell states, at a rate determined by the conservation of the quantum Fisher information.
\end{enumerate}

Since the rotations in space are in one-to-one correspondence with Cartesian reference frames,  the spin-1/2 Bell state can be regarded as a unit of Cartesian reference frame.  We call such unit a  \emph{Cartesian refbit}, borrowing a term introduced by van Enk \cite{vanenk-2005-pra,vanenk-2006-pra} in a slightly different, but closely related context.   In Section \ref{sec:fission} (Section \ref{sec:fusion}) we will study how Bell states converted into  (generated from)  Cartesian refbits in the non-asymptotic setting.   Before that, we will analyze the conversion of Bell states via probabilistic operations.

\section{Probabilistic conversions}\label{sec:probabilistic} 

In this Section we study the conditions for exact and approximate probabilistic conversions of Bell states. The problem is interesting  in view of the relation between probabilistic Bell state conversions and deterministic gate simulations, discussed in the end of the paper. 

\subsection{Exact probabilistic conversions}
 Let us start from the {\em exact} conversions, that is, the conversions that can be achieved with unit fidelity.  We focus on the $N>1$ case, because the $N=1$ case has already been treated in Subsection \ref{subsec:singlecopy}.  For $N>1$, a necessary and sufficient condition for perfect convertibility   is  the following:

\begin{theo}[Exact probabilistic conversion of angular momentum Bell states]\label{thm:perfect}
A probabilistic machine can perfectly convert $N>1$ copies of a  rotated spin-$J$ Bell state  into $M$ copies of corresponding  spin-$K$ Bell state   if and only if $NJ\ge MK$. 
\end{theo}
The proof idea is similar to the proof idea of Theorem \ref{thm:generaldet}, with the only difference that now our machine is  not constrained to operate deterministically.       Once again, we decompose the input and output states into a superpositions of states with definite values of the quantum number of the total angular momentum, as in Eq. (\ref{inputdecompj}).   To fix ideas, consider the case where both $NJ$ and $MK$ are integers.   In this case, the angular momentum number has integer values from $0$   to $NJ$ for the input state, and from $0$ to $MK$ for the output state.    If $NJ$ is larger than $MK$, the input state contains a larger set of values.   Then, we can construct a perfect probabilistic machine that filters out  the states with values of the angular momentum  larger than $MK$ and uses  the states with angular momentum between $0$ and $MK$ as  ingredients to reproduce exactly the $M$-copy output state.   Note that this machine is intrinsically probabilistic, because it has to project the input state into a subspace, and also because it has to reshape the relative weights of the terms in the quantum superposition  (\ref{inputdecompj}).  The details can be found  in  \ref{app:perfect}.  

The proof that the condition $NJ  \le MK$ is necessary for a perfect probabilistic conversion is also provided in \ref{app:perfect}.    
The proof idea is nicely linked with the impossibility of cloning quantum states.  Basically, we prove that a perfect Bell state conversion with $NJ<MK$ would allow us to perfectly convert $2NJ$ copies of a spin-$1/2$ Bell state into $2MK$ copies  of the same state, in violation of the no-cloning theorem.  



Theorem \ref{thm:perfect} tells us the maximum number of  spin-$K$ Bell states  that can be extracted \emph{perfectly} from $N$ copies of a spin-$J$ Bell  state.     As long as we insist on having no error,  the ratio between the output and input copies must satisfy the bound  
\begin{eqnarray}\label{fracjk}  \frac MN  \le    \frac J  K   \, ,
\end{eqnarray}  no matter how small  is the  probability of success.    

  In the following we will see that tolerating a small error allows one  to   achieve  much better scaling, with $M$ growing {\em quadratically}, instead of linearly with $N$.

\subsection{Asymptotic  probabilistic  conversions}

In the limit of large $N$, the performance of the probabilistic Bell state conversion is  determined by the following Theorem:  

\begin{theo}[Asymptotic probabilistic Bell state  conversion]\label{thm:Gprob}
$N$ copies of a rotated  spin-$J$ Bell state can  be probabilistically converted into $M $ copies of the corresponding  spin-$K$ Bell state with arbitrarily small error whenever $NJ$ is large compared to $\sqrt {M  K(K+1)}$. 
Conversely, every machine with  $\sqrt {M  K(K+1)}    \gg  N J$   must have non-vanishing error. 
\end{theo}
To understand the idea of the proof, it is useful to recall that the input  and output states can be decomposed into superpositions of states with different values of the total angular momentum, as in Eq. (\ref{inputdecompj}). The weights in the superposition are proportional to a Gaussian distribution with standard deviation equal to $\sqrt {N J (J+1)/3} $ for the input state, and to $\sqrt{MK(K+1)/3}$  for the output state.  On the other hand, the support of the input distribution reaches the value $NJ$.  Hence, we can modify the weights in the input state in such a way that they look like the weights in the output state, for all the values of the angular momentum until  first $c\sqrt{MK(K+1)/3}$,  where $c$ is a  constant. In this way, we obtain a state that is identical to the desired output state for all values of the angular momentum within  $c$ standard deviations.   By choosing $c$ large enough, we can make the fidelity as large as we want.  Summarizing,  the condition $NJ\gg  \sqrt {M  K(K+1)}$ guarantees a probabilistic conversion with high fidelity.   The full proof is provided in  \ref{app:asymptoprob}, where we also   show that the condition $M\gg N^2$ leads to unavoidable errors.

With respect to the deterministic machine of Theorem \ref{thm:generaldet}, the probabilistic machine  of Theorem \ref{thm:Gprob} boosts the  number of output copies from $O(N)$ to $O(N^2)$.  As it often happens for probabilistic machines \cite{chiribella-yang-2013-natcomm,combes-christopher-2014-pra,chiribella2017optimal}, the  performance enhancement comes at the price of a damped   probability of success.   For the probabilistic machine used in Theorem \ref{thm:Gprob}, the probability of success can be upper bounded in terms of $N$, $J$, and the  ratio $R  =  M/N$.   
Whenever the ratio exceeds the critical value  
\begin{eqnarray}
  R_*  :=    \frac {J (J+1) }{K (K+1)}   \, ,
\end{eqnarray}
 the probability of success is upper bounded as 
\begin{eqnarray}\label{Pgen}
 p_{\rm succ}&\le  \left(\frac {R}{R^*}\right)^{\frac 32}  ~e^{-\frac{3NJ}{2(J+1)}  \left(1-  \frac {R_*}{R}\right)}  \, .  \qquad
\end{eqnarray}
Hence, every ratio $R \ge R_*$ leads to an exponentially vanishing probability of success.  
 In other words, every violation of the conservation of the  quantum Fisher information is exponentially suppressed in the large $N$ limit.

 Due to the exponentially vanishing probability,  the probabilistic conversions are not practically relevant in the asymptotic scenario. However, they are conceptually important, because they determine the extreme boundary of what is possible in quantum mechanics. Moreover, they are important as a technical tool for studying the the simulation of rotation gates:  in the end of the paper, we will show that the fidelity of the probabilistic Bell state conversion gives upper and lower bounds on the fidelity of the {\em deterministic} gate simulation.

\section{Quantum analysers of rotational information}\label{sec:fission}
In this section we  design  machines that break down  Bell states into Cartesian refbits.  These  machines  will be called   \emph{quantum analysers}. 

\subsection{Single Bell states are unbreakable}\label{subsec:fission1}

We start from the problem of breaking down a single Bell state into units of rotational  information.   Here we show that,  no matter how large is $J$, there is no way to convert the information carried by a single spin-$J$  Bell state  into Cartesian refbits. 
    Quantitatively,   we have the following
\begin{prop}\label{prop:unbreakable}
No machine can break down a single  spin-$J$ Bell state into Cartesian refbits with fidelity larger than 
 $ 1/2  \, \left[  1+   1/(2J)  \right]$.  
 \end{prop}
Proposition \ref{prop:unbreakable} applies to both the deterministic and probabilistic machines.  It shows that the fidelity of the quantum analyzer is never equal to 1, except   in the trivial case where $J$ is already equal to $1/2$.  For every other value of $J$, the fidelity is upper bounded by $75\%$ and converges to $50\%$ in the large $J$ limit.  

Some insight into the physical origin of this result can be obtained by thinking of the spin-$J$  system  as a system of $2J$ spin-1/2 particles, constrained to the symmetric subspace.       By discarding all particles but one,  we can transform the original spin-$J$ system into a spin-$1/2$ system.    However,  this procedure will not transform a spin-$J$ Bell state into a spin-1/2 Bell state: instead, it will generate a noisy Bell state.  The bigger the total spin, the larger the noise will be.  In this picture, the physical  reason why Bell states are unbreakable is the intra-particle entanglement among the $2J$ particles constituting the spin-$J$ system.      The complete proof can be found in  \ref{app:N=1}.  
 
Proposition \ref{prop:unbreakable} shows that the reference frame information contained in the Bell states is unbreakable. The fact highlights a fundamental difference between Cartesian reference frames and reference frames for individual directions.  Consider the   \emph{spin coherent states} \cite{radcliffe-1971-jpa,arecchi-1972-pra}, namely the states defined by  
\begin{eqnarray*}   |J,J\>_g   : =  U_{g,J}    \,  |J,J\> \, ,\end{eqnarray*}
where $  |J,J\>$ denotes the eigenstate of the $z$ component of the angular momentum operator for the eigenvalue $J$.  Among the states of a single spin-$J$ system, the spin coherent states are known to be the best carriers of  information about a single direction \cite{book-holevo-probabilistic}.    Spin coherent states can be  perfectly broken down into elementary units:  indeed, it is immediate to see that there exists a quantum channel  transforming the spin coherent state $|J,J\>_g$ into $2J$  exact copies of the spin coherent state $|1/2,1/2\>_g$.  In summary,  the information about a   \emph{single} direction can be broken down into elementary units, while the information about a full Cartesian frame cannot.

\subsection{Unlocking the refbits}

Consider now the problem of breaking down  $N$ copies   of a spin-$J$ Bell state. Already for $N=2$, interesting phenomena occur.      For example,   a deterministic machine can transform two copies of a spin-$J$ Bell state into $O(J^2)$ refbits, with a fidelity of $85.6\%$ in the large $J$ limit (\ref{app:qinheping}).       Moreover,  one can also construct  a probabilistic machine that achieves unit fidelity in the large $J$ limit.   In general, we have the following Proposition: 
\begin{prop}\label{prop:break2}
There exists a probabilistic machine that transforms $N$ copies of a rotated spin-$J$ state into $M$ Cartesian refbits with fidelity 
\begin{eqnarray}
 F^{\rm prob}_{\rm Bell}  \left [    |\Phi_{g,J}\>^{\otimes N}\to |\Phi_{g,1/2}\>^{\otimes M}\right]
&\ge1-(M+1)\exp\left[-\frac{2N^2J^2}{M+1}\right] \, .  \label{boundexp}
\end{eqnarray}
\end{prop} 
The  physical origin of the result is the same as in Theorem \ref{thm:Gprob}. The quantum number of the total angular momentum goes from $0$ to $NJ$ for the input state, and each of these values has a non-zero weight.     On the other hand,  the target output state has a Gaussian distribution with variance $  O(\sqrt{ M})$. Hence,  the input state can be turned into a good approximation of the output state whenever $NJ$ is large compared to $ \sqrt{ M}$.   The bound (\ref{boundexp})   follows from Hoeffding's bound on the tails of the Gaussian distribution.      The explicit derivation is provided in \ref{app:break2}. 
       
Proposition \ref{prop:break2} tells us  that the error vanishes whenever the condition $M\ll N^2 J^2 $ is satisfied.   In short,  a probabilistic machine can  ``unlock"  the elementary units of reference frame information contained in the Bell state, whenever  $N$ is larger than 1 and the product   $N  J$   is sufficiently large.

\section{Quantum synthesizers of rotational information}\label{sec:fusion}

In this section we consider the task of generating Bell states from elementary units of rotational information.     Machines implementing this task will be called \emph{quantum synthesisers}. 

Specifically, we  study how $N$  Cartesian refbits can be converted into a single spin-$K$ Bell state.  For this task, we consider a simple protocol based on  estimation and re-reparation: given $N$ Cartesian refbits, the protocol is to    estimate the rotation and to  prepare the corresponding spin-$K$ Bell state.       Let us denote by $\hat g$ the estimate of the unknown rotation and let us assume that the machine prepares the Bell state $|\Phi_{\hat{g},K}\>$ corresponding to the estimate.   The  action of the machine will be described by the measure-and-prepare channel 
\begin{eqnarray}\label{MPchannel}
\map{C}_{\rm MP}(\rho):=\int\d\hat{g} ~|\Phi_{\hat{g},K}\>\<\Phi_{\hat{g},K}|~\Tr\left[ M_{\hat{g}} \,   \rho \right]
\end{eqnarray}
where $\{M_{\hat{g}}\}$  are the operators describing the measurement. 
 We choose the optimal measurement for the estimation of $g$.  Such measurement is  given by  the operators    \cite{chiribella2011group,chiribella-dariano-2005-pra}
\begin{eqnarray}\label{POVM}
M_{\hat{g}}=\sum_{k,l=0}^{N/2}(2k+1)(2l+1)  ~ |\Phi_{\hat{g},k}\>\<\Phi_{\hat{g},l}|  \, .
\end{eqnarray}

Now, it is interesting to ask how fast can $K$ grow as a function of $N$.  By explicit evaluation, we find out that the fidelity converges to 1 whenever $K$ grows slower than   $\sqrt N$.     In this case,  the fidelity has the asymptotic expression
\begin{equation}
F^{\rm MP}_{\rm Bell}          \Big[    | \Phi_{g,1/2}  \>^{\otimes N}  \to   | \Phi_{g,K}\>    \Big]
=1-\frac{(2K+1)^2}{4N}+O(N^{-1}) \, ,
\end{equation}
 derived in \ref{app:mp}. 
Instead,         when $K$ is large compared to  $\sqrt N$, the fidelity vanishes as $N/K^2$ in the asymptotic limit.

\section{Simulating  rotation gates}\label{sec:gate}

Our results on the conversion of Bell states have an application to the study of quantum machines that use rotations on a given system to simulate rotations on another system.   An interesting example is that of machines that use  qubit rotations to simulate rotations of higher angular momenta.     More generally, the problem is to simulate a unitary gate through the use of another gate.    Previous works on this type of simulation included the cloning of unitary gates \cite{chiribella-dariano-2008-prl,dur2015deterministic,chiribella2015universal} and other manipulations, such as inversion, charge conjugation, and controlization  of unitary gates \cite{chiribella2016optimal}.


\subsection{The gate simulation task}
Suppose that we are given a black box implementing a unitary gate $U_x$, where the parameter $x$ is randomly drawn from some set $\set X$   with probability $p_x$.    Our goal is to implement another unitary gate $V_x$, possibly acting on a different system, using $U_x$ as a resource.  The problem is how to simulate the gate $V_x$ while actually using the gate $U_x$.    For example, $U_x$ could be a rotation on a small system and $V_x$ could be a rotation on a larger system. 
 
\begin{figure}[b]
\centering
      \includegraphics[width=0.8\textwidth]{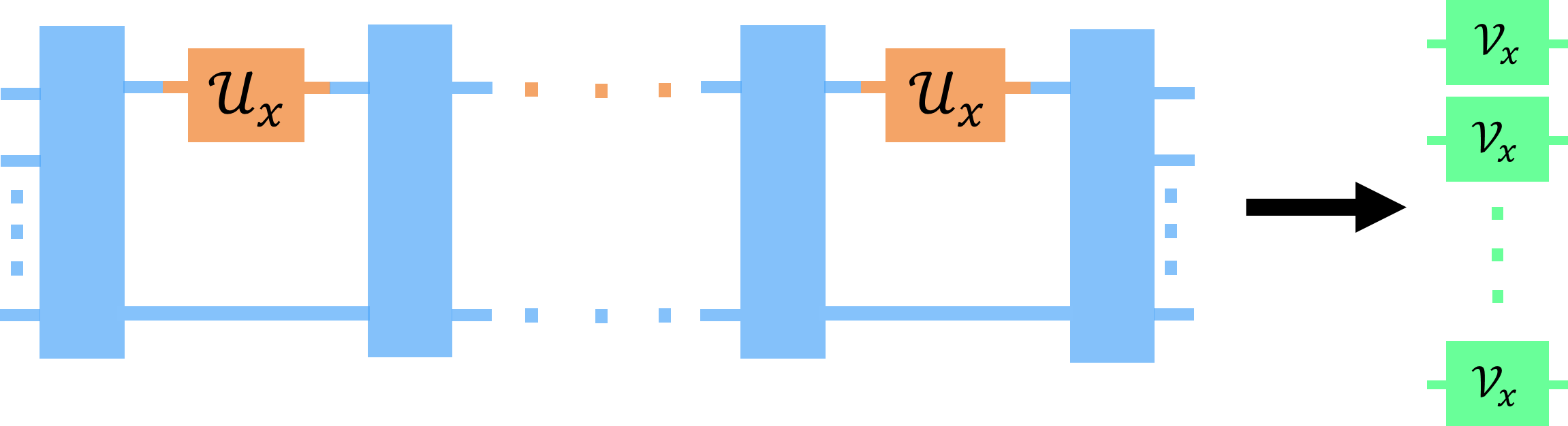}\caption{{\bf Network for gate simulation.}   
       The network  (in blue) has $N$ open slots  where $N$ uses of the unitary channel  $\map U_{x}$ (in  orange) can be inserted.  When the gates are in place,  the network simulates $M$ uses of the  unitary channel $\map V_x$ (in green). }
       \label{fig:gate}
\end{figure}

Now, suppose that  we can use the gate $U_x$ for    $N$ times  and we want to simulate $M$ parallel uses of the gate $V_x$.    To do this, we will need to build a quantum network where the black box implementing  the gate  $U_x$ is connected with other  quantum devices, suitably chosen to optimize the simulation.  A network of this kind is shown in Figure \ref{fig:gate}.    

  We will first consider the case where the gate simulation network consists of deterministic devices.  In this case, the overall operation implemented by the network is a quantum channel (trace-preserving completely positive map)   $\map{C}^{(N)}_x$,  acting on $M$ identical systems.    Ideally, the action of  $\map{C}^{(N)}_x$  should resemble as much as possible the action of $M$ parallel queries to the gate $V_{x}$.  To quantify the  resemblance, we use   the \emph{entanglement  fidelity}  \cite{horodecki-horodecki-1999-pra}, namely the fidelity between the output of the actual channel and the output of the target channel when the two channels are applied locally to a maximally entangled state.    
   
 Specifically, let $|\Phi_{\rm out}\>$ be the canonical Bell state defined by  
 \begin{eqnarray*}
 |\Phi_{\rm out}\>   =   \frac{\sum_{n=1}^{d_{\rm out}}  \,   |n\>\otimes |n\>}{\sqrt {d_{\rm out}}} \, ,
 \end{eqnarray*}
 where $d_{\rm out}$ is the dimension of the Hilbert space $\spc H_{\rm out}$, on which the target gate $V_x$ acts.    When the channel $\map C_x^{(N)}$ is applied locally on $M$ copies of the  Bell state $|\Phi_{\rm out}\>$, it generates the output state    
 \begin{eqnarray*}
\Sigma^{({\rm   det})}_x  =   \left(\map{C}^{(N)}_x\otimes \map{I}_{\rm out}^{\otimes M}\right)   \left(  |\Phi_{\rm out}\>\<\Phi_{\rm out}|^{\otimes M} \right) \, ,
 \end{eqnarray*}
 where we implicitly understand that    the channel $\map C^{(N)}_x$ (respectively,  $\map I_{\rm out}^{\otimes M}$) acts on the first (respectively, second) system of each Bell pair inside the round bracket. 
    For a fixed value of  the parameter $x$,  the entanglement  fidelity is 
  \begin{eqnarray*}   
\fl\qquad F_{  x}^{\rm det}   \Big[   (N,  U_x)    \to  V_x^{\otimes M} \Big]  =   \<\Phi|^{\otimes M} \,   (V_x^\dag\otimes  I_{\rm out})^{ \otimes M}       \,  \Sigma_x^{(\rm det)}   \,     (V_x\otimes  I_{\rm out})^{\otimes M}    |\Phi\>^{\otimes M}  \, ,
\end{eqnarray*}
 where the notation $(N,U_x)$  means that the input resource consists of $N$ uses of the gate $U_x$, employed in an arbitrary   (not necessarily parallel) disposition.      Averaging 
over all possible rotation gates, we obtain the  fidelity 
\begin{eqnarray}\label{gate}
F^{\rm det}_{\rm gate}  \Big [   (N,U _x)    \to  V_x^{\otimes M}  \Big] =\sum_{x\in\set X }  \,  p_x   ~   F_x^{\rm det}  \Big [   (N, U_x)    \to  V_x^{\otimes M}  \Big] \,. 
\end{eqnarray}
It is worth mentioning that the maximisation of the entanglement fidelity is equivalent to the maximisation of  the average fidelity between the outputs when the channels are applied to a randomly drawn input state \cite{horodecki-horodecki-1999-pra}.

We will also consider  networks of probabilistic devices, whose successful functioning is heralded by a sequence of measurement outcomes.   A probabilistic network will   transform the $N$ input uses of the gate $U_{x}$ into a (generally trace non-increasing) quantum operation $\map Q^{(N)}_x$ acting on $M$  identical systems. 
  The  probability that the quantum operation $\map Q^{(N)}_x$ takes place on the Bell state $  |\Phi_{\rm out}\>^{\otimes M}$ is  
   \begin{eqnarray*}
p ({\rm succ  }|  x) = \Tr [     (\map Q^{(N)}_x\otimes  \map I_{\rm out}^{\otimes M}) \, (  |  \Phi_{\rm out}  \>\<\Phi_{\rm out}   |^{\otimes M}    ) ]  \, ,
  \end{eqnarray*}  
  where we implicitly understand that  the quantum operation $\map Q^{(N)}_x$ acts on the first system of each Bell pair. 
When the quantum operation takes place, the
   output state   is  
   \begin{eqnarray*}
 \Sigma^{(N)}_x       =        \frac{   (\map Q^{(N)}_x\otimes  \map I_{\rm out}^{\otimes M}) \, (   |  \Phi_{\rm out}  \>\<\Phi_{\rm out}   |^{\otimes M}    ) \, }{\Tr [     (\map Q_x\otimes  \map I_{\rm out}^{\otimes M}) \, (  |  \Phi_{\rm out}  \>\<\Phi_{\rm out}   |^{\otimes M}     )]}  \, .
  \end{eqnarray*}    
For a given value of the parameter $x$, the entanglement  fidelity is 
\begin{eqnarray}
\fl \qquad F_x^{\rm prob}    \Big[    (N, U_x)  \to  V_x^{\otimes M}  \Big]   &= \<  \Phi  |^{\otimes M}   (V_x\otimes  I_{\rm out})^{\dag \,  \otimes M}    \,  \Sigma^{(N)}_x  \,    (V_x\otimes  I_{\rm out})^{\otimes M}     |  \Phi\>^{\otimes M}  \, .
\end{eqnarray}
Conditioning on the successful functioning of the devices in the network, the average fidelity is
 \begin{eqnarray}
  \label{probgate}      F_{\rm gate}^{\rm prob}    \Big[  (N,  U_x)   \to  V_x^{\otimes M}   \Big]       =    \sum_{x\in\set X}         p(x|{\rm succ}  ) \,   F_x^{\rm prob}  \Big[  (N,  U_x  )  \to  V_x^{\otimes M} \Big]\, . \qquad   
  \end{eqnarray}   

In the following we will establish connections between the fidelities of gate simulation and the fidelities of Bell state conversion. 
\subsection{Simulation of gates vs conversion of states}
  
 The Choi isomorphism sets up a one-to-one correspondence between unitary gates and maximally entangled states, whereby the gate $U$ is mapped into the state 
 \begin{equation}\label{connection}
 |\Phi_U\>  :  =      (U\otimes I)  \,   | \Phi\>\,  , \qquad  |\Phi\>   :  =  \frac 1  {\sqrt d}\sum_{n=1}^d   \,  |n\>\otimes |n\>   \,.
 \end{equation}  
Operationally, the map $U  \mapsto  |\Phi_U\>$ can be implemented deterministically by applying the gate on one system of an Bell pair.  Instead,   the inverse map $|\Phi_U\>  \mapsto U$ can only be  implemented probabilistically via conclusive teleportation  \cite{bennett1993teleporting}, with a maximum probability of success  determined directly by the causality principle  \cite{genkina2012optimal}.  

The above  properties of the Choi isomorphism imply an elementary relation between the task of simulating gates and the task of transforming Bell states:
   \begin{prop}\label{prop:trivial}
 Let $\{  U_x\}_{x\in\set X}$ and $\{  V_x\}_{x\in\set X}$ be two sets of  unitary gates and  let $\{|\Phi_{U_x}\>  \}_{x\in\set X}$  and  $\{|\Phi_{V_x}\>  \}_{x\in\set X}$ be the corresponding sets of Bell states. Then, one has   
 \begin{eqnarray}
\nonumber  F^{\rm det}_{\rm gate} \Big [ (N, U_x)  \to    V_x^{\otimes M}  \Big]   & \le      F^{\rm prob}_{\rm gate} \Big[  (N, U_x)   \to  V_x^{\otimes M} \Big] \\
&   = F^{\rm prob}_{\rm Bell}  \Big [|\Phi_{U_x}\>^{\otimes N}  \to |\Phi_{V_x}\>^{\otimes M}     \Big ]   \, ,
 \end{eqnarray} 
 where $F^{\rm prob}_{\rm Bell}  \Big[ |\Phi_{U_x}\>^{\otimes N}  \to |\Phi_{V_x}\>^{\otimes M}   \Big]$ is the optimal fidelity for the probabilistic Bell state conversion  $|\Phi_{U_x}\>^{\otimes N}  \to |\Phi_{V_x}\>^{\otimes M}  $.   \end{prop} 
    
 \medskip 
 
The above proposition is quite generic, for  it simply follows from the operational properties of the Choi isomorphism.   
More   interesting features arise when the unitaries $\{  U_x\}$ and $\{  V_x\}$ form two group representations.  These features will be discussed in the remaining part of the paper.  

\subsection{Analytical expression of the fidelity}
Let us consider first the case where a single use of the gate $U_x$ is available, corresponding to the case $N=1$.  
For simplicity of notation, we also assume that the goal is to simulate a single use of the gate $V_x$, although everything we will do holds also for $M  >1$ uses, upon replacing   $V_x$ with $V_x'  :  =  V_x^{\otimes M}$. In the following we assume that the two sets of gates  $\{  U_x\}$ and $\{  V_x\}$ are two representation of the same group $\grp G$, and we write $g \in \grp G$ in place of $x\in \set X$.   

With this notation, we have the following
\begin{theo}\label{theo:probabilisticfidelity}
For two group representations  $\{ U_g\}$ and $\{V_g\}$, one has 
\begin{eqnarray}
 \nonumber F^{\rm prob}_{\rm gate}  \Big[ U_g \to V_g\Big]   &= F^{\rm prob}_{\rm Bell}  \Big[  |\Phi_{U_g}\>  \to |\Phi_{V_g}\>    \Big] \\
  \label{morte}  &    =     \max_{l \in  {\rm Irr}  (  V \otimes \overline U)}  \,   \left[ \frac 1 {d_{\rm out} \, d_l} \,  \left (  \sum_{j \in   {\rm Irr} (  U)}  {  d_j  \, m^{(j)}_l}\right) \right]  \, ,
\end{eqnarray}
where  the maximum is over the irreducible representations contained in the decomposition  of the product representation $  \{V_g\otimes \overline U_g\}$,   $d_{\rm out}$ is the dimension of the Hilbert space where $V_g$ acts, $d_l$   is the dimension   of the  irreducible representation labelled  by $l$, the sum inside the round brackets  is over the irreducible representations contained in the decomposition of $\{  U_g\}$, and $m_l^{(j)}$ is the multiplicity of the representation $\{  U_g^{(l)}\}$ in the decomposition of $\{  V_g \otimes  \overline  U^{(j)}_g\}$.  
\end{theo}

Quite naturally, the fidelity depends only on group-theoretic quantities. These quantities are related to the structure of the input and output representations, and to the way  these representations  are combined together.  The exact value of the fidelity is derived in  \ref{app:probabilisticfidelity}.  

The probabilistic fidelity (\ref{morte}) takes an even simpler expression when the input gates $\{ U_g\}$ form an irreducible representation.    In this case, the sum over $j$ consists of a single term and one is left with the expression  
\begin{eqnarray}\label{mortenera}
\fl \quad F^{\rm prob}_{\rm gate}    \Big[  U_g  \to  V_g  \Big]= F^{\rm prob}_{\rm Bell}  \Big[ |\Phi_{U_g}\>^{\otimes N}  \to |\Phi_{V_g}\>^{\otimes M}    \Big]    =   \frac{d_{\rm in}}{d_{\rm out}} \,    \left[  \max_{l \in  {\rm Irr}  (  V \otimes \overline U)}  \,    \frac{m_l}{ d_l}   \right]  \, ,
\end{eqnarray}
where $m_l$ is the multiplicity of the representation $\{  U_g^{(l)}\}$ in the decomposition of $\{  V_g \otimes \overline U_g\}$. 
  As an illustration, consider the following example:  
  
\begin{eg}[Cloning an unknown unitary gate]\label{eg:cloning}
  Imagine that  an experimenter is given access to a single use of an unknown unitary gate  $U$, acting on a  $d$-dimensional  quantum system.   Imagine that the experimenter wants to simulate two uses of the same gate $U$.    If probabilistic operations are allowed,  the fidelity is given by Eq. (\ref{mortenera}). To evaluate the minimum over $l$, one has to decompose   the representation $\{  U\otimes U \otimes \overline U\}$, which is easily done using the machinery of Young diagrams.  Specifically, one finds that the maximum ratio $m_l/d_l$ is obtained by choosing the representation $\{  U\}$, which has dimension $d_l  =  d$ and multiplicity $m_l = 2$.   Hence, the probabilistic fidelity has the expression
    \begin{equation}\label{clonbell}
F^{\rm prob}_{\rm gate}  \Big[  U \to  U^{\otimes 2}\Big]   =   F^{\rm prob}_{\rm Bell}  \Big [  |\Phi_{U}\>  \to  |\Phi_U\>^{\otimes 2} \Big]  = \frac{2}{d^2} \, .
 \end{equation} 
 The probabilistic fidelity is an upper bound to the deterministic fidelity,   which has the value    \cite{chiribella-dariano-2008-prl}
 \begin{eqnarray}
 F^{\rm det}_{\rm gate}    \Big[  U \to  U^{\otimes 2} \Big]   =    \frac { 1  +  \sqrt  { 1  -   \frac 1{d^2}}}{d^2} \, .       
 \end{eqnarray}
 Comparing the two fidelities, we observe that the advantage of using  probabilistic operations vanishes when the dimension of the system is large: the gap between the deterministic and probabilistic fidelities vanishes as $1/d^4$. 
  \end{eg}

\subsection{No probabilistic advantage for irreducible representations} 
When the input and output representations are irreducible, it turns out that there is no difference between the performances of probabilistic and deterministic strategies. 
More precisely,  one has the following 
\begin{theo}\label{theo:prob=det}
Let $\set G$ be a group and let $\{U_g\} $ and $\{V_g\} $ be two unitary representations of $\grp G$.    
If the input representation $\{  U_g\}$ is irreducible, then one has 
\begin{eqnarray}
F^{\rm det}_{\rm Bell}  \Big [    U_g \to V_g\Big]       =  F^{\rm prob}_{\rm Bell}  \Big[   |\Phi_{U_g}\> \to |\Phi_{V_g}\> \Big]      \, .
\end{eqnarray} 
If both the input representation  $\{  U_g\}$ and the output representation   $\{  V_g\}$ are irreducible, then 
one has 
\begin{eqnarray}
F^{\rm det}_{\rm gate}    \Big [    U_g \to V_g\Big]     =  F^{\rm prob}_{\rm gate}      \Big[   |\Phi_{U_g}\> \to |\Phi_{V_g}\> \Big]    \, .
\end{eqnarray} 
\end{theo}  
The proof idea comes from the symmetry of the problem.  The key observation is that the optimal probabilistic operations can be chosen to be   invariant under the action of the gates $\{U_g\}$ and $\{V_g\}$.   Irreducibility  ensures that the probability that the operations take place is independent on the input state. In turn, independence on the input state means that each operation is proportional to a deterministic operation, which takes place with unit probability on every state.    The difference between Bell state conversions and gate simulations is only that the Bell state conversion involves operations on input systems acted upon by the representation $\{  U_g\}$, while the gate simulation involves also operations where the input state can be acted upon by  the representation  $\{  V_g\}$.    The details of the proof are provided in  \ref{app:prob=det}.

 \begin{eg}[Optimal cloning of Bell states]  
 Suppose that we are given one copy of a generic Bell state $|\Phi_U\>  $ of two $d$-dimensional quantum systems, and that   we want to generate one more copy.    This problem is to find the physical process that implements the Bell state conversion $|\Phi_U\>  \to |\Phi_U\>^{\otimes 2}$  with maximum fidelity. 
 
 The problem of cloning Bell states was previously studied in terms of single-copy fidelity   \cite{karpov2005cloning}.   Thanks to Theorem \ref{theo:prob=det}, we now know that the optimal two-copy fidelity is 
 \begin{eqnarray}
 F_{\rm Bell}^{\rm det}  \Big [ \, |\Phi_U\>  \to |\Phi_U\>^{\otimes 2} \Big]  =  \frac 2{d^2} \, .
 \end{eqnarray} 
cf.  Eq. (\ref{clonbell}). 
  \end{eg}  
  
  \begin{eg}[Optimal charge-conjugation]  
  Suppose that we are given a black box implementing the gate $U$ and we want to use it to simulate the gate $\overline U$, obtained from $U$ through complex conjugation in a fixed basis.   In physics, $\overline U$ is  sometimes regarded as the result of  charge-conjugation.  
Using Theorem \ref{theo:prob=det}, we know  that the optimal deterministic network performs equally well as the optimal probabilistic network, whose fidelity is given by Eq. (\ref{mortenera}).     The evaluation of the fidelity is simple:  one has only to decompose the representation $\{  U\otimes U\}$, which is known to have only two irreducible subspaces, the symmetric subspace and the antisymmetric subspace, of dimensions $d_s  = d (d+1)/2$ and $d_a=  d(d-1)/2$, respectively.  
  The evaluation of Eq. (\ref{mortenera}) then yields the fidelity 
  \begin{eqnarray}
   F_{\rm gate}^{\rm det}  \Big[  U\to \overline U\Big]   =    \frac 2 { d (d-1)}  \, ,    
  \end{eqnarray}
retrieving a result of Ref. \cite{chiribella2016optimal}.     Note that the fidelity is equal to 1 for two-dimensional systems, where the gates $U$ and $\overline U$ are unitarily equivalent.    
   \end{eg}
 \subsection{Local and memoryless operations}
 
\begin{figure}[b]
\centering
      \includegraphics[width=0.6\textwidth]{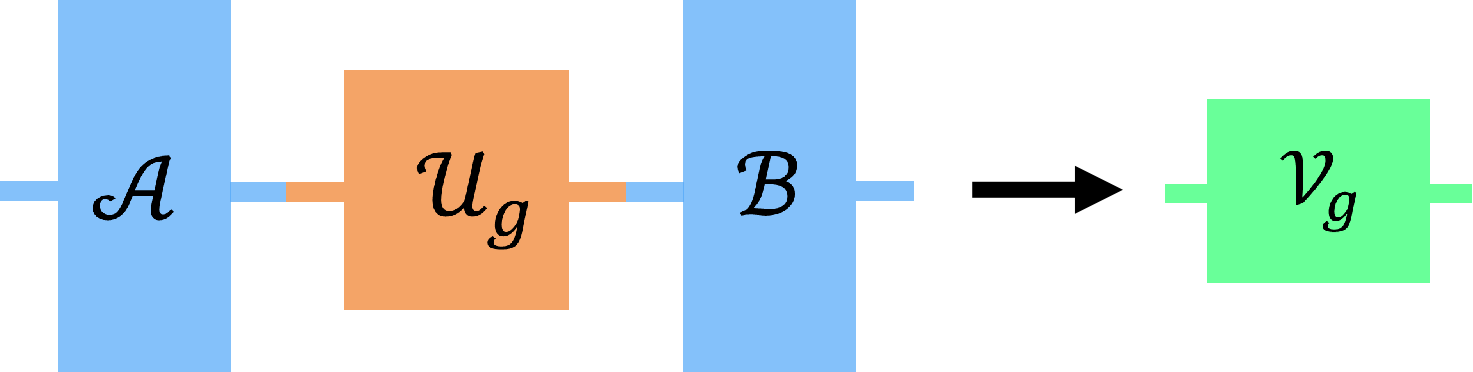}\caption{{\bf Gate simulation network without internal memories.}   
       The network  (in blue)  exploits one use    of the  unitary channel  $\map U_{g}$ (in  orange) to  simulate one  use of the  unitary channel  $\map V_g$ (in green)  without the assistance of any internal memory. }
       \label{fig:memoryless}
\end{figure}

 We conclude our analysis of the $N=1$ case with  a sufficient condition for the realization of the optimal Bell state conversion with local operations, and for the realization of the optimal gate simulation through a network without internal memories, as in Figure \ref{fig:memoryless}.  
 
 \begin{theo}\label{theo:local}
 Let $\set G$ be a group, let $\{U_g\} $ and $\{V_g\} $ be two unitary representations of $\grp G$, and let $m_l$ be the multiplicity of the irreducible representation $\{  U_g^{(l)}\}$ in the decomposition of $\{  V_g \otimes \overline U_g\}$.    If  the maximum of $m_l/d_l$ is attained by a representation with multiplicity $m_l  =1$, then
 \begin{enumerate}
 \item   the  Bell state conversion $  |\Phi_{U_g}\>  \to  |\Phi_{V_g}\>$ can be achieved by local operations whenever $\{  U_g\}$ is irreducible.     
 \item    the gate simulation $U_g \to V_g$ can be achieved by a network without internal memories whenever $\{  U_g\}$ and $\{  V_g\}$  are irreducible
 \end{enumerate}
 \end{theo}

A good illustration of all the features shown so far is the simulation of a rotation on a spin-$K$ system using a rotation on a spin-$J$ system.
    
\begin{figure}[t!]
\centering
      \includegraphics[width=0.8\textwidth]{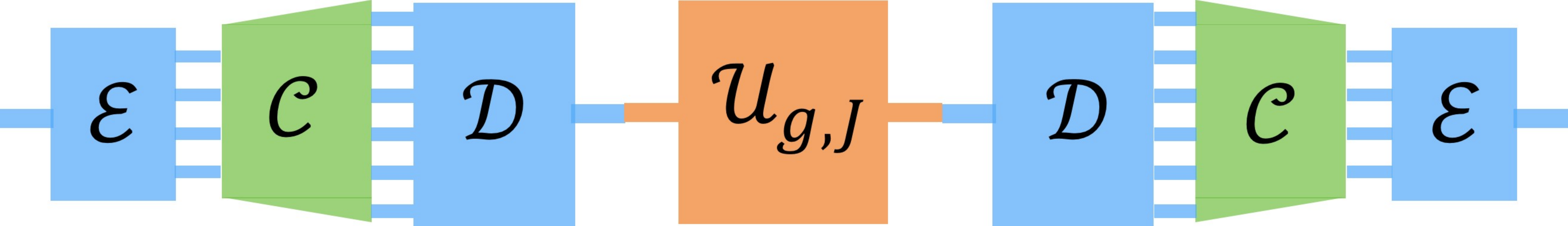}\caption{{\bf   Optimal single-use rotation converter.}   The figure illustrates the simulation of a single use of a spin-$K$ rotation gate $U_{g,K}$ using a spin-$J$ rotation $U_{g,J}$ for the $J\ge K$ case. A sequence of operations are performed on any input system of $U_{g,K}$ to change its spin: first, the encoding channel $\map{E}$ embeds the spin-$K$ system into a system of $2K$ spin-1/2 systems (qubits).   Then,    the universal cloning machine  $\map{C}$  optimally turns $2K$ qubits into $2J$ qubits.  Finally, the decoding channel  $\map{D}$  merges $2J$ qubits into a single spin-$J$ system. The system's spin now fits the spin-$J$ rotation. After applying $U_{g,J}$, the same sequence of operations is performed in reverse order and input-output spins.  A similar sequence of operations allows us to achieve conversions with  $K>J$, the only difference being that on has to replace  the universal cloning  with a universal discarding, corresponding to the partial trace over $2K-2J$ qubits. }
       \label{fig:singlegate}
\end{figure}

\begin{eg}[Rotate one spin to rotate another]
  Suppose that we have access to a gate that rotates a spin-$J$ system, can we use it to rotate a spin-$K$ system with $K\not =  J$?  
Intuitively, one would expect that the answer is affirmative, as long as  $J$ is smaller than $K$: after all, if we are able to rotate a bigger system, we should also be able to rotate a smaller one.    But this is not the case: the   entanglement fidelity of the best gate simulation is given by Eq. (\ref{mortenera}), which here gives
\begin{eqnarray}
F_{\rm gate}   \Big[U_{g,J}\to  U_{g,K}\Big]    =  \frac{  2J+1 }{  (2K+1) \,  (  2  |J-K|  + 1)}  \, ,  
\end{eqnarray}
retrieving the result of Ref. \cite{bisio-dariano-2014-pla}.  
  Except in the trivial case where $J$ and $K$ are equal or where $K$ is zero, the fidelity is always bounded away from 1, even in the asymptotic limit of large $J$ and $K$.    Specifically, one can easily see that the fidelity is upper bounded by $75\%$ for all values of $J$ and $K$  with $J\not =  K$ and  $K\not =  0$.    This  upper bound also implies an upper bound  for the simulation of multiple uses of the same rotation: a single use of the rotation $U_{g,J}$ cannot simulate     $M$ uses of the rotation $U_{g, K}$ with more than $75\%$ fidelity.  

Besides the value of the fidelity,  it is interesting to see how the optimal gate simulation is achieved.  In Fig. \ref{fig:singlegate} we show an explicit quantum circuit attaining the maximum fidelity. 
Again, the idea is to encode the state of a single spin-$J$ system into the state of $2J$ spin-1/2 particles and to use cloning and discarding in order to force the number of particles to  have the desired  values.  

\end{eg}  

\subsection{Bounding the gate fidelity in terms of the Bell state fidelity}
We conclude the paper with a fundamental result linking gate simulation and Bell state conversion for arbitrary groups: 
 \begin{theo}\label{theo:nontrivial}
Let $\set G$ be a group and let $\{U_g\} $ and $\{V_g\} $ be two finite-dimensional  unitary representations of $\grp G$. 
Then  the fidelities for the gate  simulation  $U_g\to V_g$ and for the state conversion $ |\Phi_{U_g}\> \to |\Phi_{V_g}\>$ satisfy the bound 
 \begin{equation}\label{bounds}
 F^{\rm det}_{\rm gate}  \Big [  U_g\to V_g \Big]\ge  \left(  F^{\rm prob}_{\rm Bell}   \Big [  |\Phi_{U_g}\> \to |\Phi_{V_g}\> \Big] \right)^2  \, .
  \end{equation} 
  \end{theo}
The proof is provided in \ref{app:gate}. Theorem \ref{theo:nontrivial} has an important  consequence:  for unitaries forming a group representation,  
 a gate simulation can be achieved deterministically with high entanglement fidelity \emph{if and only if} the corresponding state  conversion can be achieved probabilistically with high fidelity. This fact follows  from the bound   \begin{eqnarray}\fl \qquad \left(   F^{\rm prob}_{\rm Bell}   \Big [  |\Phi_{U_g}\> \to |\Phi_{V_g}\> \Big]    \right)^2  \le  F^{\rm det}_{\rm gate}    \Big [  U_g\to V_g \Big] \le  F^{\rm prob}_{\rm Bell}  \Big [  |\Phi_{U_g}\> \to |\Phi_{V_g}\> \Big]  \, ,\end{eqnarray}   
 implied by Theorem \ref{theo:nontrivial} and Proposition  \ref{prop:trivial}. 

  Leveraging on the correspondence between gate simulation and state conversion, we can directly derive a number of facts about the simulation of rotation gates:
  \begin{enumerate}
\item  For large $J$, two uses of  the same rotation on a spin $J$ system can be used  to deterministically simulate $O(J^2)$ uses of the same rotation on spin $1/2$ systems. The quadratic factor comes from the fact that the  simulation performance is evaluated  {\em on average} over all states.  Our result means that  $O(J^2)$ uses of the spin $1/2$ rotation can be simulated with vanishing error on the {\em typical} input states, whose total probability tends to one  in the large $J$ limit.   However, there exist states where the simulation does not work: for example, all  states in subspaces with total angular momentum of size $J^2$.  

 \item   In the large $N$ limit, $N$ uses of  a spin-$J$ rotation can be used to  deterministically simulate $   O(N^2)$ uses of the corresponding spin-$K$ rotation with exponentially vanishing error in the large  $N$ limit.    This result is reminiscent of previous results on the super-replication of quantum gates \cite{dur2015deterministic,chiribella2015universal,chiribella2016quantum}, where the task was to simulate multiple uses of a gate using a smaller number of queries to the {\em same} gate.  
     \end{enumerate}
These two examples are just an illustration of the power of Theorem \ref{theo:nontrivial}. The theorem can be applied not only to  rotation gates, but also  to every other group of gates, including  phase gates \cite{buvzek1999optimal}, permutation gates \cite{von2004quantum}, and the set of all unitary gates \cite{kahn2007fast},

\section{Conclusions}\label{sec:discussion} 
  
We showed that Bell states of different angular momenta can be reversibly converted into one another  at a rate fixed by the Fisher information.     The reversibility of the conversion means that we  can regard the Bell state of two spin-1/2 particles as the {\em Cartesian refbit}, the elementary unit of information about rotations in space.

Not every state can be converted into Cartesian refbits, though. 
   States that do  not carry faithful information about Cartesian reference frames  cannot be converted into  Cartesian refbits, even if asymptotically many  copies are available  \cite{marvian-spekkens-2014-pra-asymmetry}.     
  This fact can be easily seen for spin coherent states: since a spin coherent state encodes only one direction, having many identical copies  will   not help identifying  the other two directions needed to specify  a  full Cartesian frame.     This observation opens up two directions for future research. The first direction is to study the convertibility problem for  states that are in one-to-one correspondence with Cartesian reference frames, such as the optimal states introduced in Refs. \cite{chiribella-dariano-2004-prl,bagan-baig-2004-pra,chiribella-dariano-2005-pra,hayashi-2006-pla}.   
  Having good carriers of directional information is important not only for quantum metrology, but also for the realization of  programmable quantum processors \cite{nielsen-1997-prl,buvzek-2006-qip} that   perform  rotations in space \cite{bartlett-rudolph-2009-njp,bisio-chiribella-2010-pra,chiribella2011group} or  carry out  measurements  in a desired  basis \cite{marvian-2012-arxiv,ahmadi-jennings-rudolph-2013-njp}.  
  
   In this scenario, it is meaningful to establish the optimal asymptotic rate for the conversion of a given state  Cartesian refbits, and the optimal asymptotic rate for the inverse process.  
    We  call these two  rates the \emph{distillable refbits} and the \emph{refbits of formation}, respectively, in analogy to the corresponding notions in the resource theory of  entanglement \cite{bennett-bernstein-1996-pra}.      For angular momentum Bell states, we have shown that the distillable refbits coincide with the refbits of formation,  because every angular momentum Bell state can be reversibly converted into refbits.   Whether the equality holds for all faithful carriers of Cartesian reference frames is a genuinely open question.   Note that, in principle, the equality between distillable refbits and refbits of formation could hold even if the conversion is not implemented by reversible operations. 
 
The second  direction  is to consider quantum states that carry only partial information about rotations---for example, spin coherent states.  Expanding the scope of the ideas discussed in this paper, we propose to  adopt  the spin-1/2 coherent states as units of directional information, or \emph{directional refbits}.      The choice is motivated by the fact that \emph{i)} the spin-1/2 coherent states  are the optimal states of the smallest quantum system carrying directional information, and \emph{ii)}  spin-$J$ coherent states can be reversibly converted into spin-$1/2$ coherent states.      An interesting question is whether there exists a canonical asymptotic decomposition of all quantum states into directional refbits.
If true, this result would lead to a dramatic simplification of the resource theory of asymmetry.  


The techniques developed in this work can be extended from rotations in space to other groups of operations, including translations in time, charge conjugation, and general unitary evolutions in finite dimensions. The key open  question is whether every theory of asymmetry admits a notion of elementary unit, in which every resource state can be asymptotically decomposed.


\medskip
{{\bf Acknowledgments.} 
This work is supported by the   Hong Kong Research Grant Council through Grant No. 17326616, by National Science Foundation of China
through Grant No. 11675136, by the HKU Seed Funding for Basic Research,   by the Foundational Questions Institute (FQXi-RFP3-1325), and by the Canadian Institute for Advanced Research (CIFAR).   }

\bigskip

\bibliographystyle{unsrt}
\bibliography{ref}


\appendix

\section{Single-copy conversions}\label{app:singlecopy}

Here we provide the derivation of our results on single-copy conversions.  Specifically, we  show that  
\begin{enumerate}  
\item  {\em The optimal deterministic and probabilistic machine perform equally well in the single-copy conversion.}  This is because the set of input states  
 \begin{eqnarray*}      \{   |\Phi_{g,J}  \>    =   ( U_{g, J}\otimes I) \,  |\Phi_J\>  ~|~   g\in   \grp {SO} (3) \}\end{eqnarray*}
 is invariant under the action of the irreducible representation $\{  U_{h,J} \otimes  \overline U_{k,J} ~|~  h \in   \grp {SO} (3)  \, ,   k  \in   \grp {SO} (3)    \}$.     It is a general fact that probabilistic machines do not offer any advantage whenever the set of input states is invariant under an irreducible representation \cite{chiribella-yang-2013-natcomm}.  
 
\item   {\em The optimal fidelity of the single-copy Bell state conversion is given by Eq. (\ref{JFK}).}    The proof of Eq. (\ref{JFK}) follows from a  general expression of the probabilistic fidelity of single-copy Bell state conversions,  derived in  Eq. (\ref{mortenera}) of Section \ref{sec:gate}, and summarized here for convenience.    The expression applies to Bell states of the form $  |\Phi_{U_g}\>  :  =  ( U_g\otimes I_{\rm in}) |\Phi_{\rm in}\>$  and $|\Phi_{V_g}\> :  =  (  V_g  \otimes I_{\rm out})  |\Phi_{\rm out}\>$, where  $\{  U_g\}$ and $\{  V_g\}$ are two representations of a given group $\grp G$,  and $  |\Phi_{\rm in}\>  $    (respectively, $|\Phi_{\rm out}\>$)  is the canonical Bell state in the Hilbert space $  \spc  H_{  \rm in} \otimes \spc   H_{\rm in  }$  (respectively, $  \spc  H_{  \rm out} \otimes \spc   H_{\rm out  }$), on which the representation $\{  U_g\}$  (respectively, $\{  V_g\}$) acts locally.  With these settings, the optimal probabilistic fidelity is 
   \begin{eqnarray}\label{duplicato}
   F^{\rm prob}_{\rm Bell}  \Big[ |\Phi_{U_g}\>  \to |\Phi_{V_g}\>    \Big]    =   \frac{d_{\rm in}}{d_{\rm out}} \,    \left[  \max_{l \in  {\rm Irr}  (  V \otimes \overline U)}  \,    \frac{m_l}{ d_l}   \right]  \, ,
\end{eqnarray}
where $m_l$ is the multiplicity of the irreducible representation $\{  U_g^{(l)}\}$ in the decomposition of the product representation $\{  V_g \otimes \overline U_g\}$  (we direct the interested reader to  the textbook \cite{book-fulton-harris-representation}  for more details on the  notions of irreducible representation and multiplicity).       Here we are interested in the case where the group is  $\grp G  =  \grp{SO}  (3)$ and the  representations are    $U_g   \equiv  U_{g,  J}$ and $V_g \equiv U_{g,  K}$.  With these settings, the addition rules for the angular momenta  imply  that the product  representation $  \{V_g\otimes \overline U_g\}$ is decomposed into representations with angular momentum    $l$ running from $|J-K|$ to $J+K$.  The dimension of such representations is  $d_l  =  2l+1$, while the multiplicity is  $m_l= 1$ for every $l$.   Hence, the maximum in Eq. (\ref{duplicato}) is attained when $l=  |J-K|$, thus implying  
\[     F^{\rm prob}_{\rm Bell}  \Big[ |\Phi_{g,J}\>  \to |\Phi_{g,K}\>    \Big]   =  \frac  {  2J+1}{  (2K+1)  (2  |J-K|  +1)} \, ,\]
as stated by  Eq.  (\ref{JFK}).     

\item {\em The protocol described in Figure \ref{fig:clontrace} is  optimal.}  The protocol is based on local operations that convert the input spin-$J$ systems into output spin-$K$ systems.  Each spin-$J$  (spin-$K$) system is regarded as a composite system of $2J$   ($2K$) spin-$1/2$ particles, whose state is constrained to  the symmetric subspace.   
The conversion  of the $2J$ input  particles into $2K$ output particles  is implemented by the quantum channel $\map C_{2J\to 2K}$,  defined by the relation
\begin{eqnarray*}
\map{C}_{2J\to 2K}  (\rho):=\left\{\begin{array}{ll} \left(\frac{2J+1}{2K+1}\right) \, P_{2K}  \left(\rho\otimes I^{\otimes 2(K-J) } \right) \, P_{2K} & J\le K\\ \Tr_{J-K}[\rho]  & J>K \end{array}\right.
\end{eqnarray*}
where  $P_{2K}$ is the projector on the symmetric subspace of $2K$  spin-$1/2$ particles, while $\rho$ is a generic state of the input system.

Applying the channel $\map C_{2J \to 2K}$ on each of the two spin-$J$ systems of the input  Bell state $|\Phi_{g,J}\>$, we obtain the output state 
\begin{eqnarray*}
\fl\rho^{\rm (out)}_{g}  =   \frac{2J+1}{(2K+1)^2}\sum_{j,j'=-J}^{J} & \sum_{k,k'=-K+J+j}^{K-J+j} \frac{{2J\choose{J+j}}{2J\choose{J+j'}}{2K-2J\choose{K-J+k-j}}{2K-2J\choose{K-J+k'-j}}}{\sqrt{{2K\choose K+k}{2K\choose K+k'}{2K\choose K+k+j'-j}{2K\choose K+k'+j'-j}}}\\
&\times|K,k\>_g\<K,k+j'-j|_g\otimes|K,k'\>_g\<K,k'+j'-j|_g
\end{eqnarray*}
 for $J\le K$,  or the output state
\begin{eqnarray*}
\fl\rho^{\rm (out)}_{g}  =     \frac{1}{2J+1}\sum_{j,j'=-J}^{J} &\sum_{k,k'=K-J+j}^{-K+J+j} \frac{{2J-2K\choose{J-K+j-k}}{2J-2K\choose{J-K+j-k'}}\sqrt{{2K\choose K+k}{2K\choose K+k'}{2K\choose K+k+j'-j}{2K\choose K+k'+j'-j}}}{{2J\choose{J+j}}{2J\choose{J+j'}}}\\
&\times|K,k\>_g\<K,k+j'-j|_g\otimes|K,k'\>_g\<K,k'+j'-j|_g
\end{eqnarray*}
for $J>K$. Here we defined 
\[  |K,k\>_g:=U_{g,K}|K,k\>\]   and we used the expression 
\[\fl \quad \sqrt{{2M\choose M+m}}|M,m\>_g=\sum_{n}\sqrt{{2N\choose N+n}{2M-2N\choose M-N+m-n}}|N,n\>_g|M-N,m-n\>_g\] 
valid for  $M\ge N$.

Using the expressions of the output state $\rho_g^{(\rm out)}$, we can now compute the fidelity of the Bell state conversion. 
 For  $J  \le K$, we obtain 
 \begin{eqnarray*}
\fl \qquad F_{\rm Bell}    \Big[   |\Phi_{g,J}\>\to |\Phi_{g,K}\>  \Big ]&  =\int\d g\,\<\Phi_{g,K}|\rho^{\rm (out)}_{g}|\Phi_{g,K}\>\\
&=\frac{2J+1}{(2K+1)^3}\sum_{k,j,j'}\frac{{2J\choose J+j}{2J\choose J+j'}{2K-2J\choose K-J+k-j}^2}{{2K\choose K+k}{2K\choose K+k+j'-j}}\\
&=\frac{2J+1}{(2K+1)^3}\sum_{k,j}\frac{{2J\choose J+j}{2K-2J\choose K-J+k-j}}{{2K\choose K+k}}\sum_{j'}\frac{{2J\choose J+j'}{2K-2J\choose K-J+k-j}}{{2K\choose K+k+j'-j}}\\
&=\frac{2J+1}{(2K+1)^3}\sum_{k,j}\frac{{2J\choose J+j}{2K-2J\choose K-J+k-j}}{{2K\choose K+k}}\cdot\frac{2K+1}{2K-2J+1}\\
&=\frac{2J+1}{(2K+1)(2K-2J+1)}.
\end{eqnarray*}
For $J>K$, we obtain 
\begin{eqnarray*}
\fl F_{\rm Bell}\Big[   |\Phi_{g,J}\>\to |\Phi_{g,K}\>  \Big ]&=\frac{1}{(2J+1)(2K+1)}\sum_{k,j,j'}\frac{{2J-2K\choose{J-K+j-k}}^2{2K\choose K+k}{2K\choose K+k+j'-j}}{{2J\choose{J+j}}{2J\choose{J+j'}}}\\
&=\frac{1}{(2J+1)(2K+1)}\sum_{k,j}\frac{{2J-2K\choose{J-K+j-k}}{2K\choose K+k}}{{2J\choose{J+j}}}\sum_{j'}\frac{{2K\choose K+k+j'-j}{2J-2K\choose{J-K+j-k}}}{{2J\choose{J+j'}}}\\
&=\frac{1}{(2J+1)(2K+1)}\sum_{k,j}\frac{{2J-2K\choose{J-K+j-k}}{2K\choose K+k}}{{2J\choose{J+j}}}\cdot\frac{2J+1}{2J-2K+1}\\
&=\frac{2J+1}{(2K+1)(2J-2K+1)}.
\end{eqnarray*}
In both cases, the fidelity is equal to the optimal fidelity in Eq. (\ref{JFK}).
   \end{enumerate}

\section{Proof of Eq. (\ref{inputdecompj}): decomposition of the  input Bell states }\label{app:inputdecomp}


\Proof All throughout the paper we will make extensive use of the  the \emph{double-ket notation} \cite{royer-1991-pra,dariano-presti-2000-pla}, which associates operators with bipartite states according to the correspondence $A   \to  |A\kk$, where $|A \kk $ is the bipartite state defined as
\begin{eqnarray*}|A\kk:=\sum_{m,n}A_{mn} \, |m\>\otimes |n\>\, , \qquad A_{mn}  :=   \<  m  |   A  |n\>  \,.   \end{eqnarray*} 
Using this  notation, the rotated Bell states can be expressed as  
\begin{eqnarray*}  |\Phi_{g,J}\>   =  \frac{  | U_{g,J}  \kk}{\sqrt{2j+1}} \, .\end{eqnarray*}
To  deal with the $N$-copy states $|\Phi_{g,J}\>^{\otimes N}$ we take advantage of 
the    decomposition of the corresponding tensor product Hilbert space.  For simplicity, we assume $N$ and $M$ to be even (anyways,  the parity of $N$ and $M$ will  not matter in the asymptotic limit) . For each  Bell pair, we denote by  $\spc H_{J,A}$    ($\spc H_{J,  B}$)  the Hilbert space of the first  (second) spin.     Then, for $x=   A, B$ we have the decomposition 
 \begin{eqnarray}\label{hilbertspacedec}
\left(\spc H_{J,x}\right)^{\otimes N}  =  \bigoplus_{j=0}^{N  J}  \,   \left( \spc R_{j,  x}  \otimes \spc M^{(N,J)}_{j,x} \right)  \, , 
\end{eqnarray}
where $j$ is the quantum number of the total angular momentum,  $ \spc R_{j,x}$  is a representation space carrying the $(2j+1)$-dimensional irrep of $\grp  {SO}  (3)$, and $\spc M^{(N,J)}_{j,x}$ is a multiplicity space, where the group $\grp  {SO}  (3)$ acts trivially.    
 Relative to this decomposition, the action of the rotation gates  $U_{g,J}^{\otimes N}$ can be expressed as 
    \begin{eqnarray}\label{unitarygatedec}
  U_{g,J}^{\otimes N}  =  \bigoplus_{j=0}^{N  J}  \,   \left( U_{g,j}  \otimes I_{m_j^{(N,J)}}   \right)  \, , 
\end{eqnarray} 
where $ \{ U_{g,j}\}$ is the irreducible representation of $\grp {SO} (3)$ with quantum number $j$, acting on the representation space $\spc R_j$, while  $  I_{ m_j^{(N,J)}}$ denotes the identity operator on the multiplicity space  $\spc M^{(N,J)}_{j}$. The dimension of the representation space  $\spc R_j$ is $d_j  =  2j+1$.  The dimension of the multiplicity space $\spc M^{(N,J)}_{j}$, denoted by $m_j^{(N,J)}$, is called the {\em multiplicity of the irreducible representation $\{  U_{g,J}\}$ in the decomposition of the product representation $\{  U_{g,J}^{\otimes N}\}$}. 
 More details on the  decomposition (\ref{unitarygatedec}), sometimes called the {\em isotypic decomposition} can be found in the classic textbook by Fulton and Harris \cite{book-fulton-harris-representation}.

Using the  decomposition (\ref{hilbertspacedec}),  the Hilbert space of $N$ Bell pairs can be decomposed as 
\begin{eqnarray*}
\nonumber   \left (  \spc H_{J,A}  \otimes     \spc H_{J,B}   \right)^{\otimes N}      &
  &\simeq    \bigoplus_{j,j'=0}^{NJ}        \left( \spc R_{j,A}   \otimes \spc R_{j',B}     \otimes \spc M^{(N,J)}_{j,A}    \otimes \spc M_{j',B}^{(N,J)} \right)   \, ,     
\end{eqnarray*} 
where we rearranged the Hilbert spaces in such a way that the representation spaces are on the left and the multiplicity spaces are on the right. 
Plugging Eq. (\ref{unitarygatedec}) into the double-ket notation, the $N$-copy input state can be represented as 
\begin{eqnarray}  
\nonumber  |\Phi_{g,J}\>^{\otimes N}  &  =    \frac  {  |U_{g,J}  \kk^{\otimes N}}{\sqrt { d_J^N}} \\
\nonumber   & =  \frac1  {  \sqrt{d_J^N} }  \,  \bigoplus_{j = 0}^{NJ}      \,    |   U_{g,j}\kk  \otimes  |  I_{m_j^{(J,N)}}  \kk      \\  
  \label{inputSW}   &   =  \bigoplus_{j = 0}^{NJ}           \sqrt{p^{(N,J)}_{j}} \left|\Psi_{g,j}^{(N,J)}\right\> \, , 
\end{eqnarray}
where   $ |\Psi_{g,j}^{(N,J)}\>$ is the state
\begin{equation}  
 \left|\Psi_{g,j}^{(N,J)}\right\>:=\frac{|U_{g,j}\kk}{\sqrt{d_j}}\otimes\frac{|I_{m^{(N,J)}_j}\kk}{\sqrt{m_{j}^{(N,J)}}}  \, ,
\end{equation}  
$d_j  :  =    \dim  \spc R_j  =  2j+1 $ and $  m_{j}^{(N,J)}  :  =  \dim  \spc M^{(N,J)}_{j,J}$ are the dimension and the multiplicity of the representation $\{  U_{g,j}\}$, respectively, and   $p^{(N,J)}_{j}$ is the probability distribution 
\begin{eqnarray}\label{pdef}
p^{(N,J)}_{j}: =\frac{ d_j\, m^{(N,J)}_{j}}{(2J+1)^{N}}  \, .
\end{eqnarray}
\qed

Note that the  decomposition of Eq. (\ref{inputdecompj}) also applies to the  output state, which takes the form 
\begin{eqnarray}
\label{outputSW}  \fl\left|\Phi_{g,K}\right\>^{\otimes M}=\bigoplus_{k  =  0 }^{M  K} \sqrt{p^{(M,K)}_{k}} \left|\Psi_{g,k}^{(M,K)}\right\> \, ,  \qquad \left|\Psi_{g,k}^{(M,K)}\right\>:=\frac{|U_{g,k}\kk}{\sqrt{d_k}}\otimes\frac{|I_{m_k^{(M,K)}}\kk}{\sqrt{m^{(M,K)}_{k}}} \, ,
\end{eqnarray}
where  $m^{(M,K)}_{k}$ is the multiplicity of the irreducible representation $\{  U_{g, k}\}$ in the decomposition of the product representation $\{  U_{g,K}^{\otimes M}\}$.  

\section{Proof of Eq. (\ref{papprox}): asymptotic expression for $p^{(N,J)}_{j}$.}\label{app:asymptotic}

\Proof  By definition, one has $p^{(N,J)}_{j}   =  d_j m^{(N,J)}_{j}/(2J+1)^N$.  
To compute the multiplicity, we use the  standard group-theoretic formula   \cite{book-fulton-harris-representation}   
\begin{eqnarray*}
m^{(N,J)}_{j}=\int \d g~\Tr\left[ \,\overline U_{g,J}\right]^{N}\Tr[ U_{g,j}] \,  ,
\end{eqnarray*}
which follows from the orthogonality of the irreducible characters  \cite{book-fulton-harris-representation}.   
Now, we parametrize the rotation $g$ in terms of the rotation angle $\omega$ and of the rotation axis $\st n$.  Integrating over all possible directions $\st n$, we obtain the expression 
\begin{eqnarray*} 
m^{(N,J)}_{j}&  = \int_{-\pi}^{\pi}\d\omega\frac{(2J+1)^{N}}{\pi}\sin\frac{(2j+1)\omega}{2}\sin\frac{\omega}{2}\left[\frac{\sin\frac{(2J+1)\omega}{2}}{(2J+1)\sin\frac{\omega}{2}}\right]^N.
\end{eqnarray*}
Since $\lim_{N\to\infty}\left[\frac{\sin(J+\frac{1}{2})\omega}{(2J+1)\sin\frac{\omega}{2}}\right]^{N}=0$ for fixed nonzero $\omega\in[-\pi,\pi]$, we can constrain the above integral to a small interval centred  around the origin $\omega=0$. Hence, we have 
\begin{eqnarray*}
\fl m^{(N,J)}_{j}\approx \int_{-\delta}^{\delta}\d\omega~\frac{(2J+1)^{N}}{\pi}\sin\frac{(2j+1)\omega}{2}\sin\frac{\omega}{2}\exp\left\{N\ln\left[\frac{\sin\frac{(2J+1)\omega}{2}}{(2J+1)\sin\frac{\omega}{2}}\right]\right\}\qquad N\gg1
\end{eqnarray*}
 where we set $\frac{1}N\ll\delta\ll1$. The Taylor expansion of $\ln\left[\frac{\sin(J+\frac{1}{2})\omega}{(2J+1)\sin\frac{\omega}{2}}\right]$ yields 
 \begin{eqnarray*}
 \ln\left[\frac{\sin(J+\frac{1}{2})\omega}{(2J+1)\sin\frac{\omega}{2}}\right]=-\frac{J(J+1)}{6}\omega^{2}+O(\omega^{4}) \, .
 \end{eqnarray*}
  Using this expansion, we can express $m^{(N,J)}_{j}$ as 
\begin{eqnarray}
\nonumber \fl m^{(N,J)}_{j} & =\frac{(2J+1)^{N}}{2\pi}  \,     \int_{-\delta}^{\delta}\d\omega\left[\cos j\omega-\cos(j+1)\omega\right]\exp\left[-\frac{NJ(J+1)}{6}\omega^{2}\right]  \\  
 \label{app1}
  \fl &  \qquad   \times \left[1+O(\delta^4)\right].
\end{eqnarray}
It is straightforward to see that, since $\delta\gg1/N$, one has
\begin{eqnarray*}
\lim_{N\to\infty}\int_{\delta}^{\infty}\d\omega\left[\cos j\omega-\cos(j+1)\omega\right]\exp\left[-\frac{NJ(J+1)}{6}\omega^{2}\right]=0.
\end{eqnarray*}
On the other hand, the same integral with the range $[-\delta,\delta]$ is non-vanishing in the large $N$ limit. As a consequence, we can expand the range of the integral in Eq. (\ref{app1}) from $[-\delta,\delta]$ to $[-\infty,\infty]$, introducing only a negligible error. Adopting this expansion,  we obtain the following approximate value of the multiplicity:
\begin{eqnarray*}
m^{(N,J)}_{j}&\approx\frac{(2J+1)^{N}}{2\pi}\int_{-\infty}^{\infty}\d\omega\left[\cos j\omega-\cos(j+1)\omega\right]\exp\left[-\frac{NJ(J+1)}{6}\omega^{2}\right]\\
&=\sqrt{\frac{3(2J+1)^{2N}}{2\pi NJ(J+1)}}\exp\left[-\frac{3j^{2}}{2NJ(J+1)}\right]\left[1-e^{-\frac{3(2j+1)}{2NJ(J+1)}}\right].
\end{eqnarray*}
By  Taylor expansion of the last term, we finally obtain the asymptotic expression 
\begin{eqnarray*}
m^{(N,J)}_{j}&=\sqrt{\frac{27(2J+1)^{2N}}{8\pi N^{3}J^{3}(J+1)^{3}}}\exp\left[-\frac{3j^{2}}{2NJ(J+1)}\right]\left[1-O\left(\frac{1}{N(J+1)}\right)\right].
\end{eqnarray*}
Substituting the above into Eq. (\ref{pdef}), we find that $p^{(N,J)}_{j}$ is given by 
\begin{eqnarray}\label{papprox'}
\fl \qquad p^{(N,J)}_{j}=\sqrt{\frac{27(2j+1)^{4}}{8\pi N^{3}J^{3}(J+1)^{3}}}\exp\left[-\frac{3j^{2}}{2NJ(J+1)}\right]\left[1-O\left(\frac{1}{N (J+1)}\right)\right] \, ,
\end{eqnarray}
as anticipated in  Eq. (\ref{papprox}). \qed

\section{Proof of Theorem \ref{thm:generaldet}: asymptotic convertibility via deterministic reversible operations}\label{app:asymptodet}

Here we provide the proof  of Theorem \ref{thm:generaldet}.  The ingredients of the proof are collected in the following subsections.

\subsection{Covariant isometric channels}\label{subsec:covariant}
As an ansatz for the Bell state conversion, we consider   covariant isometric channels, namely channels that 
\begin{enumerate}
\item satisfy the covariance condition  
\begin{eqnarray*}  \map C  \left( U_{g,J}^{\otimes N}    \, \cdot   \,        U_{g,J}^{\otimes N  \dag}  \right)    =        \left( U_{g,K}\right)^{\otimes M}      \,  \map C \left(  \, \cdot \, \right)  \,     \left( U_{g,K}\right)^{\otimes M \dag}  \,  , \forall g\in  \grp{SO} (3) \, ,\end{eqnarray*}
and
 \item  can be written as 
$\map C  (\cdot)=V\cdot V^\dag$
for some isometry $V$. 
\end{enumerate} 
Such channels are guaranteed to exist when $NJ$ and $MK$ are both integers, and $NJ$ is smaller than or equal to $MK$.  

The maximum fidelity over all covariant isometric channels is given by the following Proposition 
\begin{prop}\label{prop:isocov}
For $NJ\le MK$, the fidelity of the optimal covariant isometric channel is 
\begin{eqnarray}\label{Fdet}
F^{\rm iso}_{\rm Bell} \Big[   |\Phi_{g,J}\>^{\otimes N}\to |\Phi_{g,K}\>^{\otimes M}  \Big ]&=\left(\sum_{j=0}^{NJ}\sqrt{p^{(N,J)}_{j}p^{(M,K)}_{j}}\right)^2 \, , 
\end{eqnarray}  
where $p_j^{(N,J)}$ and $p_k^{(M,K)}$ are the probabilities in the decompositions (\ref{inputSW}) and (\ref{outputSW}), respectively.
The optimal isometric channel is defined by the relation
\begin{eqnarray}\label{channel}
V\left|\Psi_{g,j}^{(N,J)}\right\>=\left|\Psi_{g,j}^{(M,K)}\right\>   \qquad \forall j\in[0,NJ]\, , \quad \forall g\in\grp{SO} (3) \, \, ,
\end{eqnarray}
where the states   $|\Psi_{g,j}^{(N,J)} \>$ and  $|\Psi_{g,j}^{(M,K)} \>$  are defined in Eqs.   (\ref{inputSW}) and (\ref{outputSW}), respectively.
\end{prop}

\Proof 
 In terms of the isometry  $V$,   the covariance requirement amounts to the relation
\begin{eqnarray}\label{bbb}
  \left(U_{g,K}  \otimes U_{h,K}\right)^{\otimes M}V\left(U_{g,J}^{\dag}\otimes U^\dag_{h,J}\right)^{\otimes N}=V \qquad \forall g, h\in \grp{SO}(3).
  \end{eqnarray} 

Hence, the fidelity of the isometric channel $\map V  (\cdot)  =  V\cdot V^\dag$ is given by
\begin{eqnarray*}
\fl \qquad F_{\rm iso}  \Big[   |\Phi_{g,J}\>^{\otimes N}\to |\Phi_{g,K}\>^{\otimes M}  \Big ]& =  \left|     \<  \Phi_K |^{\otimes M}   ~ V |  \Phi_J\>^{\otimes N} \right|^2 \\
& =  \left|\sum_{j=0}^{\min\{NJ,MK\}}\sqrt{p_{j}^{(N,J)}p_{j}^{(M,K)}}\left\<\Psi^{(N,J)}_j\right|V\left|\Psi^{(M,K)}_j\right\>\right|^2 \\
& \le\left(\sum_{j=0}^{\min\{NJ,MK\}}\sqrt{p^{(N,J)}_{j}p_{j}^{(M,K)}}\right)^2 \, .
\end{eqnarray*}
The bound is saturated by the isometry defined in Eq. (\ref{channel}), which is therefore optimal over all covariant isometries. \qed 

An alternative optimality proof can be obtained from an upper bound on the fidelities of covariant isometric channels, derived by Marvian and Spekkens  in  Theorem 3 of Ref. \cite{marvian-spekkens-2013-njp}.

\subsection{Evaluation of the asymptotic fidelity} 

The asymptotic fidelity for the Bell state conversion  can be computed by inserting the asymptotic expression (\ref{papprox'}) into  the expression for the fidelity (\ref{Fdet}).  
Suppose that  deviation  $\Delta  :  =  MK(K+1) -  N  J (J+1)$  grows as $N^{1-\alpha}$ with   $\alpha \in  (0,1/4)$.   Then, for $NJ\le MK$ we obtain the asymptotic fidelity   
\begin{eqnarray}
 F^{\rm iso}_{\rm Bell}\Big[   |\Phi_{g,J}\>^{\otimes N}\to |\Phi_{g,K}\>^{\otimes M}  \Big ]
&=1-\frac{3\Delta^2}{8S^2}-O\left(\frac{\Delta^4}{S^4}\right)-O\left(\frac{1}{\sqrt{S}}\right)  \, ,\label{FDeet}
\end{eqnarray}
with $S:=NJ(J+1)$.   Note that the fidelity converges to $1$ in the large $N$ limit. 
     The condition $\alpha  <1/4$  can be easily removed:  if one can produce up to $N^{1-\alpha}$ extra copies with vanishing error, one can   always discard some copies and reduce the number of extra copy to $N^{1-\alpha'}$ with $\alpha' \ge  1/4$.    Note also that the condition $NJ  \le MK$, used to derive Eq. (\ref{Fdet}), can also be removed:  if  $NJ  > MK$ one can construct an isometry  from a subspace of the input space and complete the isometry with some other operation in the orthogonal subspace. With this choice, the fidelity will have at least the value of Eq. (\ref{FDeet}), meaning that the error will vanish at least as $(\Delta/  S)^2$.   This concludes the proof of Theorem \ref{thm:generaldet}. \qed

\section{Derivation of Eq. (\ref{asymptoticupper}): asymptotic upper bound on the deterministic fidelity}\label{app:det-bound}
Here we provide a bound on the fidelity of arbitrary quantum channels  in the limit of large $N$.    The  proof technique is a generalization of a technique introduced in our previous work \cite{chiribella-yang-2014-njp} for the cloning of qubit Bell states, corresponding to the $J=K=1/2$ case.

Due to the symmetry of the problem, the optimal quantum channel $\map C$ can be assumed without loss of generality to be covariant, i.e. to satisfy the condition
\begin{eqnarray}\label{covariant1}
\left(\map{U}_{g,K}\otimes\map{U}_{h,K}\right)^{\otimes M}\map{C}=\map{C}\ \left(\map{U}_{g,J}\otimes\map{U}_{h,J}\right)^{\otimes N}.
\end{eqnarray}
In terms of the Choi operator $C:=(\map{C}\otimes I)(|I\kk\bb I|)$, the covariance condition (\ref{covariant1})  can be   written as
\begin{eqnarray}\label{covariant2}
\left[C, (U_{g,K}\otimes U_{h,K})^{\otimes M}\otimes(  \overline{U}_{g,J} \otimes {\overline U}_{h,J})^{\otimes N}\right]=0.
\end{eqnarray}
Here $\overline {U}_{g,J}$ is the complex conjugation of $U_{g,J}$.

With this constraint, as well as the property of the Choi operator that $\map{C}(\rho)=\Tr_{\rm in}[(I_{\rm out}\otimes \rho^T)C]$, the fidelity (\ref{Fdef}) can be rewritten as
\begin{eqnarray*}
\fl \qquad F^{\rm det}_{\rm Bell} \Big[   |\Phi_{g,J}\>^{\otimes N}\to |\Phi_{g,K}\>^{\otimes M}  \Big ]=\Big( \<\Phi_K|^{\otimes M}\otimes \<\Phi_J|^{\otimes N}\Big)  \,  C\,   \Big  (|\Phi_J\>^{\otimes N}  \otimes |\Phi_K\>^{\otimes M}\Big)  \, .  
\end{eqnarray*}

To evaluate the fidelity, we use Eqs. (\ref{inputSW})   and (\ref{outputSW}) to decompose the joint state of the input and the output as
\begin{eqnarray}
\fl\nonumber |\Phi_J\>^{\otimes N} \otimes  |\Phi_K\>^{\otimes M} &   =   \left( \bigoplus_{j=0}^{NJ}  \,   \sqrt{p_j^{(N,J)}} \frac{ |I_j\kk}{\sqrt d_j}   \otimes  \frac{|  I_{m_j^{(N,J)}}\kk}{m_j^{(N,J)}}  \right)   \otimes  \left(     \bigoplus_{k=0}^{MK}  \, \sqrt{p_k^{(M,K)}} \,\frac{  |I_k\kk}{\sqrt{d_k}}  \otimes  \frac{  |  I_{m_k^{(M,K)}}\kk}{  \sqrt{m_k^{(M,K)}}}  \right)     \\
\nonumber &   =           \left( \bigoplus_{j=0}^{NJ}   \bigoplus_{k=0}^{MK}    \,  \sqrt{p_j^{(N,J)}   p_k^{(M,K)}}   \frac{ |I_j\kk}{\sqrt{d_j}}   \otimes \frac{  |I_k\kk}{\sqrt{d_k}}    \otimes  \frac{|  I_{m_j^{(N,J)}}\kk }{\sqrt{m_j^{(N,J)}}}   \otimes        \frac{|  I_{m_k^{(M,K)}}\kk}{\sqrt{m_k^{(M,K)}  } }  \right)  \\    
\label{statedecomp}
  &=\bigoplus_{l=0}^{NJ+MK}|I_l\kk  \otimes |\alpha_l\> \, ,
  \end{eqnarray}
 with \begin{eqnarray}
|\alpha_l\>:=\bigoplus_{j,k:(j,k)\to l}\sqrt{\frac{p^{(M,K)}_{k}p^{(N,J)}_{j}}{d_k d_j}}\frac{|I_{m_k^{(M,K)}}  \kk}{\sqrt{m^{(M,K)}_{k}}}\otimes\frac{|I_{m_j^{(N,J)}}\kk}{\sqrt{m^{(N,J)}_{j}}} \, ,
\end{eqnarray}
 $(j,k)\to l$ being a shorthand for the values of $j$ and $k$ satisfying the inequality $|k-j|\le l\le k+j$.
Applying Schur's lemma to Eq. (\ref{covariant2}), and taking into consideration the decomposition of the states (\ref{statedecomp}), the Choi operator can be assumed without loss of generality to have the  form
\begin{eqnarray}
&C=\bigoplus_{l=0}^{NJ+MK} (I_{l}\otimes I_{l}\otimes A_l)\\
&A_l = \bigoplus_{\scriptsize\begin{array}{c}j,k: (j,k)\to l,\\j',k': (j',k')\to l\end{array}}[A_l]_{(j,k)(j',k')}\frac{|I_{k}^{(M)}  \kk  \bb  I_{k'}^{(M)}|}{\sqrt{m^{(M,K)}_{k}  m^{(M,K)}_{k'}}}\otimes \frac{|I_{m_j^{(N,J)}}  \kk\bb  I_{m_{j'}^{(N,J)}}  |}{\sqrt{m_{j}^{(N,J)}m_{j'}^{(N,J)}}} \, .  \qquad  \qquad 
\end{eqnarray}

The fidelity is then bounded as
\begin{eqnarray*}
\fl F^{\rm det}_{\rm Bell}  \Big[   |\Phi_{g,J}\>^{\otimes N}\to |\Phi_{g,K}\>^{\otimes M}  \Big ]&=\sum_{l=0}^{NJ+MK} ~d_l\sum_{\scriptsize\begin{array}{c} j,k:(j,k)\to l,\\j',k': (j',k')\to l\end{array}}[A_l]_{(j,k)(j',k')}\sqrt{\frac{p^{(M,K)}_{k}p^{(N,J)}_{j}q_{k'}^{(M,K)}p_{j'}^{(N,J)}}{d_k d_j d_{k'} d_{j'}}}\\
&\le\sum_{l=0}^{NJ+MK} d_l\left(\sum_{j,k:(j,k)\to l}a_{jkl}\sqrt{\frac{p^{(M,K)}_{k}p^{(N,J)}_{j}}{d_k d_j}}\right)^2
\end{eqnarray*}
where $a_{jkl}:=\sqrt{[A_l]_{(j,k)(j,k)}}$, having used the positivity of the matrix $A_l$. 

Now the problem is to upper bound the function 
\begin{eqnarray*}
S&=\sum_{l=0}^{NJ+MK} d_ls_{l}^2\\
s_l&:=\sum_{j,k:(j,k)\to l}a_{jkl}\sqrt{\frac{p^{(M,K)}_{k}p^{(N,J)}_{j}}{d_k d_j}}.
\end{eqnarray*}
under the trace preservation constraint
\begin{eqnarray*}
\sum_{j,k:(j,k)\to j}d_{l}^2a_{jkl}^2=d_{j}^2\quad \forall j\in\{0,...,NJ\}.
\end{eqnarray*}
The method of Lagrange multipliers shows that the optimal coefficients $\{a_{jkl}\}$ satisfy
\begin{eqnarray*}
a_{jkl}=\frac{s_l}{d_l\lambda_j}\sqrt{\frac{p^{K,M}_k p_j^{J,N}}{d_k d_j}},
\end{eqnarray*}
where $\lambda_j\ge 0$ are the multipliers. Then the trace preservation condition becomes
\begin{eqnarray*}
\sum_{k,l:(k,l)\to j}\frac{s_l^2 p^{(M,K)}_k  p^{(N,J)}{j}}{d_k d_j}=\lambda_j^2 d_j^2 \quad\forall j\in\{0, \dots, NJ\}.
\end{eqnarray*}

Combining the expressions for the fidelity, for $s_l$ and for the optimal coefficients $\{a_{jkl}\}$, we then obtain
\begin{eqnarray}\label{S}
S&=\sum_{l=0}^{NJ+MK} d_ls_l\left(\sum_{ j,k:(j,k)\to l}a_{jkl}\sqrt{\frac{p_{k}^{(M,K)}p_{j}^{(N,J)} }{d_k d_j}}\right)\nonumber\\
&=\sum_{l=0}^{NJ+MK} s_l\left(\sum_{j,k:(j,k)\to l}\frac{s_l p_{k}^{(M,K)}p_{j}^{(N,J)}}{\lambda_j d_k d_j}\right)\nonumber\\
&=\sum_{j=0}^{NJ}\frac{1}{\lambda_j}\sum_{k,l:(k,l)\to j}\frac{s_l^2 p^{(M,K)}_k  p^{(N,J)}_{j}}{d_k d_j}\nonumber\\
&=\sum_{j=0}^{NJ} d_{j}^2\lambda_j.
\end{eqnarray}
Now that the upper bound of the fidelity depends only on $\lambda_j$, we continue the derivation by noticing that
\begin{eqnarray*}
s_l&=\sum_{j,k:(j,k)\to l}a_{jkl}\sqrt{\frac{p^{(M,K)}_{k}p_{j}^{(N,J)}}{d_k d_j}}\\
&=\sum_{j,k:(j,k)\to l}\frac{s_l p^{(M,K)}_{k}p^{(N,J)}_{j}}{\lambda_j d_k d_j d_l},
\end{eqnarray*}
which, for $s_l\ne 0$, implies that $\{\lambda_j\}$ are determined by the set of constraints
\begin{eqnarray}\label{constraint1}
d_l=\sum_{j,k:(j,k)\to l}\frac{p^{(M,K)}_{k}p^{(N,J)}_{j}}{\lambda_j d_k d_j}.
\end{eqnarray}

Notice that for any $\lambda_j\not=0$ there exists at least one $s_l\not=0$, which we define as $l(j)$, such that $\lambda_j$ appears in the $l$-th constraint. Defining the set $\set{H}_l:=\{j|l(j)=l\}$ we turn the constraints (\ref{constraint1}) into
\begin{eqnarray}\label{constraint2}
d_l=\sum_{j\in \set{H}_l}\frac{p^{(N,J)}_{j}}{\lambda_j  d_j}\sum_{k:(j,k)\to l}\frac{p^{(M,K)}_{k}}{d_k}.
\end{eqnarray}

Again, we optimize $S$ under the set of constraints (\ref{constraint2}) using Lagrangian multipliers, Eq. (\ref{S}) yielding
\begin{eqnarray*}
S&\le\sum_{l=0}^{NJ+MK}\frac{1}{d_l}\left(\sum_{j\in\set{H}_l}\sqrt{p^{(N,J)}{j}d_j\sum_{k:(j,k)\to l}\frac{p^{(M,K)}_{k}}{d_k}}\right)^2\\
&\le\left(\max_{k  \in \{0, \dots, MK\} }   \frac{p^{(M,K)}_{k}}{d_k^2}\right)\sum_{l=0}^{NJ+MK}\frac{1}{d_l}\left(\sum_{j\in\set{H}_l}\sqrt{p^{(N,J)}_{j}d_j\sum_{k:(j,k)\to l}d_k}\right)^2\\
&\le\left(\max_{k  \in \{0, \dots,  MK\}}  \frac{p^{(M,K)}_{k}}{d_k^2}\right)\sum_{l=0}^{NJ+MK}\left(\sum_{j\in\set{H}_l}\sqrt{p^{(N,J)}_{j}}d_j\right)^2\\
&\le\left(\max_{k  =  \{0,\dots,  MK\}} \frac{p^{(M,K)}_{k}}{d_k^2}\right)\left(\sum_{j\in \set H_l} \sqrt{p^{(N,J)}_{j}}d_j\right)^2
\end{eqnarray*}
having used the inequality $d_l d_j\ge\sum_{k:(j,k)\to l}d_k$ in the third inequality.
Finally, the fidelity is upper bounded as
\begin{eqnarray*}
\fl F^{\rm det}_{\rm Bell}  \Big[   |\Phi_{g,J}\>^{\otimes N}\to |\Phi_{g,K}\>^{\otimes M}  \Big ]&\le\left(\max_{k\in[0,MK]} \frac{p^{(M,K)}_{k}}{d_k^2}\right)\left(\sum_{j=0}^{NJ}\sqrt{p^{(N,J)}_{j}}d_j\right)^2\\
&=\sqrt{\frac{27}{8\pi M^3K^3(K+1)^3}}  \\
  & \quad \times\sqrt{\frac{27}{8\pi N^3J^3(J+1)^3}}\left[\int_0^{NJ}\d j ~(2j+1)^2~e^{-\frac{3j^2}{2NJ(J+1)}}\right]^2\\
&=\left[\frac{NJ(J+1)}{MK(K+1)}\right]^{\frac32}+O\left(\sqrt{\frac{N}{M^3}}\right).
\end{eqnarray*}
One can immediately see from the above bound that the deterministic fidelity vanishes in the asymptotic limit if $M\gg N$.

\section{Proof of Theorem \ref{thm:perfect}: necessary and sufficient condition for perfect probabilistic conversion }\label{app:perfect}
\subsection{The proof of sufficiency}\label{subapp:sufficiency}
The sufficiency of the condition  $NJ\ge MK$  (with $N>1$)   can be proved by straightforwardly.  If $NJ$ and $MK$ have the same parity, 
we use  the pure quantum operation $\map{M}(\rho)=W\rho W^\dag$, where $W$ is the operator defined by the relation
\begin{eqnarray}\label{Myes}
W\left|\Psi_{g,j}^{(N,J)}\right\>=\sqrt{\left(\min_k \frac{p^{(N,J)}_{k}}{p^{(M,K)}_{k}}\right)\frac{p^{(M,K)}_{j}}{p^{(N,J)}_{j}}}\left|\Psi_{g,j}^{(M,K)}\right\>  \, , 
\end{eqnarray}
required to be valid for all $j$ that $p_{N,j}\not=0$ and for every $g\in\grp{SU}(2)$.
For $N   >1$, the fidelity  of this quantum operation  can be derived by substituting the expression of $\map M$   into Eq. (\ref{Fdef-prob}). The result is 
\begin{eqnarray}\label{Fprob}
F^{\rm prob}_{\rm Bell} \Big[   |\Phi_{g,J}\>^{\otimes N}\to |\Phi_{g,K}\>^{\otimes M}  \Big ] =    \left\{  
\begin{array}{ll} 
  1   &  MK  \le NJ   \\ \\
   \sum_{k=0}^{NJ}   \,  q^{(M,K)}_{k}  \qquad &  MK  >  NJ      \,  .
  \end{array}  \right.  
\end{eqnarray}
This proves that the condition $MK\le NJ $ guarantees a perfect probabilistic conversion when $NJ $ and $MK$ have the same parity.  

Let us consider now  the case where  $NJ$ and $MK$ does not have the same parity.  If $MK$ is smaller than $NJ$, a perfect conversion can be accomplished by the following protocol: 
\begin{enumerate}
\item First analyze the spin-$J$ Bell states into $2NJ$ copies of spin-$1/2$ Bell states;
\item discard one copy of the spin-$1/2$ Bell state;
\item transform the remaining states into $M$ spin-$K$ Bell states.
\end{enumerate}
The transformations in (i) and (iii) can be accomplished perfectly, using the machine (\ref{Myes}).


\subsection{Proof  of necessity.}

Here we show that no probabilistic machine can achieve perfect conversion when $NJ<MK$. The idea is that, if such a machine existed, it would violate  the probabilistic version of the no-cloning theorem \cite{duan-guo-1998-prl}.

Let us first consider the case when $M>1$. Suppose that we are given $2NJ$ copies of an unknown spin-$1/2$ Bell state $|\Phi_{g,1/2}\>$. Then,   the sufficient condition in \ref{subapp:sufficiency} guarantees that  we can probabilistically convert the $2NJ$  spin-$1/2$ Bell states into $N$ copies of the spin-$J$ Bell state $|\Phi_{g,J}\>$, without any error.  At this point, we  can apply the machine $\map{C}$, getting $M$ copies of $|\Phi_{g,K}\>$.    But then,   the sufficient condition  in \ref{subapp:sufficiency} would imply that we can generate $2MK$ perfect copies of the state $|\Phi_{g,1/2}\>$. The overall process is a perfect cloning of the spin-$1/2$ Bell state $|\Phi_{g,1/2}\>$ since $MK>NJ$. This contradicts with the fact that only states drawn from a set of linearly independent states can be perfectly cloned, using probabilistic machines \cite{duan-guo-1998-prl}.

Finally we consider the case when $M=1$. If there exists such a machine that perfectly implements the conversion $|\Phi_{g,J}\>^{\otimes N}\to|\Phi_{g,K}\>$, then the conversion $|\Phi_{g,J}\>^{\otimes 2N}\to|\Phi_{g,K}\>^{\otimes 2}$ can also be perfectly implemented by using the machine twice. Applying again the previous argument we reach the contradiction.

\section{Proof of theorem \ref{thm:Gprob}: asymptotic probabilistic Bell state conversions}\label{app:asymptoprob}  

\subsection{Direct part}  

Here we show that the condition $ (NJ)^2\gg    M K (K+1)$ is sufficient for asymptotic convertibility with vanishing error.  
 To this purpose, we consider  as an ansatz the quantum operation defined in Eq. (\ref{Myes}). The fidelity of this particular operation, already evaluated in Eq. (\ref{Fprob}),  takes the form 

\begin{eqnarray}
F^{\rm prob}_{\rm Bell}  \Big[   |\Phi_{g,J}\>^{\otimes N}\to |\Phi_{g,K}\>^{\otimes M}  \Big ]=  \sum_{k=0}^{NJ}   \,  p^{(M,K)}_{k}  
\end{eqnarray}
when $MK$ is larger than $NJ$.    In this case, one has the bound
\begin{eqnarray*}
F^{\rm prob}_{\rm Bell}  \Big[   |\Phi_{g,J}\>^{\otimes N}\to |\Phi_{g,K}\>^{\otimes M}  \Big ]&= 1-\sum_{k=NJ+1}^{MK}   p^{(M,K)}_{k}\\
&\ge 1-3M \, {\rm erfc}\left[\sqrt{\frac{3N^2J^2}{2MK(K+1)}}~\right]\\
&\ge 1-\sqrt{\frac{6M^3K(K+1)}{\pi N^2 J^2}}\exp\left[-\frac{3N^2J^2}{2MK(K+1)}\right],
\end{eqnarray*}
where ${\rm erfc}(x):=\left(2/\sqrt{\pi}\right)\int_x^{\infty}e^{-t^2}dt$ is the complementary error function.
From the above inequalities, we can see that an asymptotically faithful conversion is  achieved whenever  $(NJ)^2\gg MK(K+1)$.

\subsection{Converse part}\label{app:converse}

We now show that the condition  $(NJ)^2\gg MK(K+1)$   is necessary for Bell state conversions with asymptotically  vanishing error.   
To this purpose, we first derive an explicit formula for the optimal probabilistic fidelity.    

\begin{lemma}\label{lem:optprobfid}
The probabilistic fidelity for the  Bell state conversion  $|\Phi_{g,J}\>^{\otimes N}  \to  |\Phi_{g,K}\>^{\otimes M}$ is  
      \begin{equation}
    \fl  \label{mortina'}
      F_{\rm Bell}^{\rm prob} \Big[   |\Phi_{g,J}\>^{\otimes N}\to |\Phi_{g,K}\>^{\otimes M}  \Big ]  = 
       \max_{l  \in \{ 0, \dots,  NJ+MK\} }   \left[  \frac 1{  d_{K}^M  d_l}   \left(    \sum_{j=  \max\{  0, \ l-MK\} }^{l+MK}   \,    d_j  \,m_l^{(M,K,j)}   \right)  \right]  \, ,   
 \end{equation}   
where $m_l^{(M,K,j)}$ is the multiplicity of the  irreducible representation $\{  U_{g,l}\}$ in the decomposition of the representation $\{  U_{g,K}^{\otimes M}  \otimes \overline U_{g,j}\}$. 
\end{lemma}

\Proof The derivation is based on an  expression for the optimal fidelity \cite{chiribella-xie-2013-prl}, which here takes the form
\begin{eqnarray}\label{xie'}
\fl F_{\rm Bell}^{\rm prob} \Big[   |\Phi_{g,J}\>^{\otimes N}\to |\Phi_{g,K}\>^{\otimes M}  \Big ] =\Big\|(I_{K}^{\otimes M}\otimes  I_{K}^{\otimes M} \otimes \tau^{-\frac12})\rho\,(I_{K}^{\otimes M}\otimes I_{K}^{\otimes M} \otimes\tau^{-\frac12})\Big\|_{\infty} \, ,
\end{eqnarray}
where  $\|  \cdot\|_\infty$  is the operator norm (in this case equal to the maximum eigenvalue of the operator inside the bars),    $\tau$  is the average input state 
\[\tau=\int\d g ~|\overline \Phi_{g,J}\>\<\overline \Phi_{g,J}|^{\otimes N}  \, ,\]
with average  with respect to  the normalized Haar measure $\d g$,   and  
   $\rho$ is the average output-input state 
 \[\rho=\int \d g~|{\Phi}_{g,K}\>\<{\Phi}_{g,K}|^{\otimes M}\otimes|\Phi_{g,K}\>\<\Phi_{g,K}|^{\otimes N}  \, . \]

The state $\tau$ can be computed from the decomposition of the input state in  Eq. (\ref{inputSW}). Using Schur's lemma  \cite{book-fulton-harris-representation} we obtain  
\begin{eqnarray*}
\tau=  \, \bigoplus_{j= 0}^{NJ} \,  p_j^{(N,J)}   ~  \frac{I_{j}}{d_j}  \otimes \frac{I_{j}}{d_j}\otimes \frac{|I_{m^{(N,J)}_j}\kk\bb I_{m^{(N,J)}_j}|}{m_j^{(N,J)}} \, .
\end{eqnarray*}   The inverse square root $\tau^{-1/2}$ is then given by  
\begin{eqnarray*}
\tau^{-1/2}=  \bigoplus_{j=0}^{NJ}    \sqrt{\frac{d^2_j}{p^{(N,J)}_j}  }   ~  I_{j}\otimes I_{j}\otimes  \frac{ |I_{m^{(N,J)}_j}\kk\bb I_{m^{(N,J)}_j}|}{m^{(N,J)}_j} \, .
\end{eqnarray*}
Hence, we have the relation   
\begin{eqnarray}
\nonumber
&    I_{\rm K}^{\otimes M}  \otimes  I_{K}^{\otimes M}  \otimes  \tau^{-\frac 12} =  \\
\nonumber 
&  =   \bigoplus_{j=0}^{NJ}   \, \sqrt{\frac{d_j^2}{p_j^{(N,J)}}  } ~  I_{K}^{\otimes M} \otimes I_{K}^{\otimes M} \otimes I_{j}\otimes I_{j}\otimes     \frac{  |I_{m^{(N,J)}_j}\kk\bb I_{m^{(N,J)}_j}|}{m^{(N,J)}_j}   \\
 &  =   \bigoplus_{j=0}^{NJ}   \bigoplus_{l,l'    =   0}^{MK+j}   \,   \sqrt{\frac{d_j^2}{p_j^{(N,J)}}  }  ~  I_{l} \otimes I_{l'} \otimes I_{m_l^{(M,K,j)}}  \otimes I_{m_{l'}^{(M,K,j)}}  \otimes   \frac{ |I_{m^{(N,J)}_j}\kk\bb I_{m^{(N,J)}_j}|}{m^{(N,J)}_j} \qquad  \qquad     
  \label{I*tau'} 
\end{eqnarray}
where $m_l^{(M,K,j)}$ is the multiplicity of the representation $\{  U_{g,l}\}$ in the decomposition of the representation $\{  U_{g,K}^{\otimes M}\otimes \overline U_{g,j}\}$. 

Let us compute the average input-output state $\rho$.   To this purpose, we first decompose the input state as in Eq. (\ref{inputSW}), obtaining 
\begin{eqnarray}
\fl \nonumber |{\Phi}_{g,K}\>^{\otimes M}  \otimes|\overline\Phi_{g,J}\>^{\otimes N} &  =  \frac {  |  U_{g,
K}\kk^{\otimes M}}{\sqrt {d_K^M}} \otimes  \frac {  |  \overline U_{g,J}\kk^{\otimes N}}{\sqrt {d_J^N}}   \\
\nonumber &  =        \frac{     |  U_{g,K}\kk^{\otimes M}  }{\sqrt{  d_K^M  }}  \otimes  \left(   \bigoplus_{j= 0 }^{NJ}   \, \sqrt{p_{j}^{(N,J)}}  \,   \frac{|\overline   U^{(j)}_g\kk}{\sqrt{d_j}}  \otimes \frac{ |I_{m^{(N,J)}_j}  \kk}{\sqrt{m_j^{(N,J)}}}   \right)    \\
&  =     \bigoplus_{j= 0}^{NJ}  \bigoplus_{l=0}^{MK+j}     \,   \sqrt{  p_j^{(N,J)}  \,   p_l^{(M,K,j)}} \,  \frac{ |   U_{g,l}\kk}{\sqrt{d_l}}       \otimes   \frac{  | I_{m_l^{(M,K,j)}}   \kk  }{\sqrt{m_l^{(M,K,j)}}}  \otimes \frac{ |I_{m^{(N,J)}_j}  \kk}{\sqrt{m_j^{(N,J)}}}  \, ,  \qquad \qquad \label{equazione'}
\end{eqnarray}
where we defined the probability distribution  
\[  p_l^{(M,K,j)}  :  =\frac{  d_l  m_l^{(M,K,j)}}{d_K^M  d_j} \, .   \]
Exchanging the order of the two summations in Eq. (\ref{equazione'})
 we obtain  the expression 
 \begin{eqnarray}\label{utile}
|{\Phi}_{g,K}\>^{\otimes M}  \otimes|\overline\Phi_{g,J}\>^{\otimes N} 
&   =  \bigoplus_{l=0  }^{NJ+ MK}           ~  \frac{|  U^{(l)}_g\kk }{\sqrt{d_l}}  \otimes      |\mu_l\> \, ,
\end{eqnarray}
where the vector $|\mu_l\>$ is define as  
\begin{eqnarray}\label{mul} 
|\mu_l\>   :  =    \bigoplus_{j=\max\{0,  l-MK \}}^{l+ MK}   \,   \,    \sqrt{  p_l^{(M,K,j)} \, p_j^{(N,J)}    }  ~   \frac{  | I_{m_l^{(M,K,j)}}   \kk  }{\sqrt{m_l^{(M,K,j)}}}  \otimes \frac{ |I_{m^{(N,J)}_j}  \kk}{\sqrt{m_j^{(N,J)}}}   \, . 
\end{eqnarray}

 Using Eq. (\ref{utile}), we can now compute the average state $\rho$, which reads
\begin{eqnarray}
\nonumber  \rho  &=\int \d g~|{\Phi}_{g,K}\>\<{\Phi}_{g,K}|^{\otimes M}\otimes|\overline \Phi_{g,J}\>\<\overline \Phi_{g,J}|^{\otimes N}\\
  &  =      \bigoplus_{l=0}^{NJ+MK}           \,  \frac{  I_l}{d_l}  \otimes \frac{ I_l}{d_l}  \otimes  |\mu_l   \>\<\mu_l |      \, . 
\end{eqnarray}

Combining  this expression with Eqs.   (\ref{I*tau}) and  (\ref{rho'}), we obtain the relation
\begin{eqnarray}
\nonumber  &(I_{\rm out}\otimes  I_{\rm out} \otimes \tau^{-\frac12})\rho\,(I_{\rm out}\otimes I_{\rm out} \otimes\tau^{-\frac12})  \\
  \label{bastissima'} &  =     \bigoplus_{l= 0 }^{NJ+MK}     \,      \frac{  I_l}{d_l}  \otimes \frac{  I_l}{d_l}  \otimes   |\mu'_l\>\<\mu'_l| \, ,
\end{eqnarray}
with 
\begin{eqnarray*}
|\mu_l'\>  :  =     \bigoplus_{j=  \max\{0,  l-MK\} }^{l+MK}      \,  \sqrt{d_j^2   \,   p_l^{(M,K,j)} } \,    \frac{ | I_{m_{l}^{(M,K,j)}} \kk}{\sqrt{m_l^{(M,K,j)}}}  \otimes  \frac{ |I_{m_j^{(N,J)}}\kk}{\sqrt{m_j^{(N,J)}}}   \, .  
\end{eqnarray*}
We now reached the conclusion.  In order to compute the fidelity, Eq. (\ref{xie'}) tells us that we must compute the maximum eigenvalue of the operator in Eq. (\ref{bastissima'}).    The eigenvalues are  
\begin{eqnarray*}  \lambda_{l}    &  =    \frac{1}{ d_l^2 }  \,  \<\mu_l'|\mu_l'\>    \\
&  =  \frac1 {d_l^2}  \,  \sum_{j=  \max\{0,  l-MK\}}^{l+ MK}   d_j^2    \,    p_l^{(M,K,j)}  \\
&  =    \,  \sum_{j=  \max\{0,  l-MK\}}^{l+ MK}   \frac{d_j    \,     m_l^{(M,K,j)}}{  d_K^M \, d_l}  
\end{eqnarray*}
Maximizing over $l$, we obtain
      \begin{equation}\label{mortina''}
 \fl      F_{\rm Bell}^{\rm prob} \Big[   |\Phi_{g,J}\>^{\otimes N}\to |\Phi_{g,K}\>^{\otimes M}  \Big ] = 
       \max_{l  \in \{ 0, \dots,  NJ+MK\} }   \left[  \frac 1{  d_{K}^M  d_l}   \left(    \sum_{j=  \max\{  0, \ l-MK\} }^{l+MK}   \,    d_j  \,m_l^{(M,K,j)}   \right)  \right]  \, .   
 \end{equation}   
\qed  

Our second step is to derive an upper bound on the probabilistic fidelity. The bound is as follows: 
\begin{lemma}
For $MK  \ge NJ$, the optimal probabilistic fidelity for the Bell state conversion    $ |\Phi_{g,J}\>^{\otimes N}\to |\Phi_{g,K}\>^{\otimes M} $ is upper bounded as  
 \begin{equation}
 \fl \qquad F^{\rm prob}_{\rm Bell}   \Big[   |\Phi_{g,J}\>^{\otimes N}\to |\Phi_{g,K}\>^{\otimes M}  \Big ]   
  \le  \frac{1+P}2  \,  \qquad   P  := \max_{x\ge 0}   \left(     \sum_{k=x}^{x +  NJ}  \,  p^{(M,K)}_{k} \right)\, , 
  \end{equation}
  with the convention that $p^{(M,K)}_{k}  =  0$ if $k$ is larger than $MK$. 
\end{lemma}
 
We are now ready to prove the converse part of Theorem \ref{thm:Gprob}.  Using Eq. (\ref{mortina''}),  we obtain
\begin{eqnarray}\label{Fprob-opt} 
 F_{\rm Bell}^{\rm prob}\Big[   |\Phi_{g,J}\>^{\otimes N}\to |\Phi_{g,K}\>^{\otimes M}  \Big ] = \max_{l \in \{  0, \dots,  NJ+MK\}} \sum_{k=0}^{MK}p^{(M,K)}_{k}~ f_{k}^{(l)} 
\end{eqnarray}
where $f_k^{(l)}$ is the function defined by 
\begin{eqnarray*}
f_{k}^{(l)}:=\frac 1 {{d_k d_l}}  \, \sum_{j=|k-l|}^{\min\{NJ,k+l\}}  \,  d_j \, .
\end{eqnarray*}
Note that the coupling of angular momenta guarantees that the function  $g_{N,k}^{(l)}$ is upper bounded as 
\begin{eqnarray}
f_{k}^{(l)}  \le 1 \, ,\qquad \forall N,k,l \,. 
\end{eqnarray}
Moreover, one has the relations 
\begin{eqnarray}
f_{k}^{(l)}=0\qquad\ {\rm for}\quad l>NJ  \quad {\rm and } \quad k<l-NJ \, ,
\end{eqnarray}
\begin{eqnarray} 
f_{k}^{(l)}  \le  \frac 12  \qquad    &  {\rm for} \quad l>NJ  \quad {\rm and} \quad k>l \, ,
\end{eqnarray}
and 
\begin{eqnarray}
f_{k}^{(l)}&\le
\frac12\qquad     {\rm  for } \quad l\le NJ  \quad {\rm and} \quad  k>NJ \, .
\end{eqnarray} 
Taking the above  conditions  into account, the fidelity can be upper bounded as
\begin{eqnarray}
 \fl \qquad F^{\rm prob}_{\rm Bell}   \Big[   |\Phi_{g,J}\>^{\otimes N}\to |\Phi_{g,K}\>^{\otimes M}  \Big ] 
  &\le  \frac{1+P}2  \,  \qquad   P  := \max_{x\ge 0}   \left(     \sum_{k=x}^{x +  NJ}  \,  p^{(M,K)}_{k} \right)\, . 
\label{bound-Fprob}
\end{eqnarray}\qed

Eq. (\ref{bound-Fprob}) tells us that   the fidelity $F^{\rm prob}_{\rm Bell} \Big[   |\Phi_{g,J}\>^{\otimes N}\to |\Phi_{g,K}\>^{\otimes M}  \Big ]$ can approach 1 only if the probability $P$ approaches 1.   Now, for large $M$, the probability distribution $p^{(M,K)}_{k}$ is approximately a polynomial times a normal distribution with standard deviation $O\left(\sqrt{MK(K+1)}\right)$, cf. Eq.  (\ref{papprox'}).    Hence, the probability $P$ can approach 1 only if the size of the interval $[x, x+  NJ]$ is comparable with $\sqrt{MK(K+1)}$.  Instead, if the ratio between $NJ$ and  $\sqrt{MK(K+1)}$ tends to zero, then the probability $P$ tends to zero, too.  In that case, the fidelity tends to the constant value $F^{\rm prob}_{\rm Bell}  =  1/2$. 
  This concludes the proof of Theorem \ref{thm:Gprob}.  \qed
  
\section{Upper bound on the fidelity of quantum analyzers with  $N=1$}\label{app:N=1}
 For machines taking a single copy as  input, we have already seen in \ref{app:singlecopy}  that deterministic and probabilistic machines have the same optimal  fidelity.  
  
Now, recall that the single-copy fidelity is an upper bound to the global fidelity, because one can always discard all the output copies but one, thus obtaining a machine that produces a single output copy.  Hence, we obtain the bound 
\begin{eqnarray*}
F_{\rm Bell }   \Big[    | \Phi_{g,J}  \>  \to   | \Phi_{g,1/2}\>^{\otimes M}    \Big] \le \frac{2J+1}{4J}  \qquad \forall  M\in\N\, , 
\end{eqnarray*}
which follows  from inserting the value $  K=  1/2$ into Eq. (\ref{JFK}).

\section{Deterministic analyzer for two copies of a spin-$J$ Bell state}\label{app:qinheping}

\begin{figure}[h!]
\centering
      \includegraphics[width=0.6\textwidth]{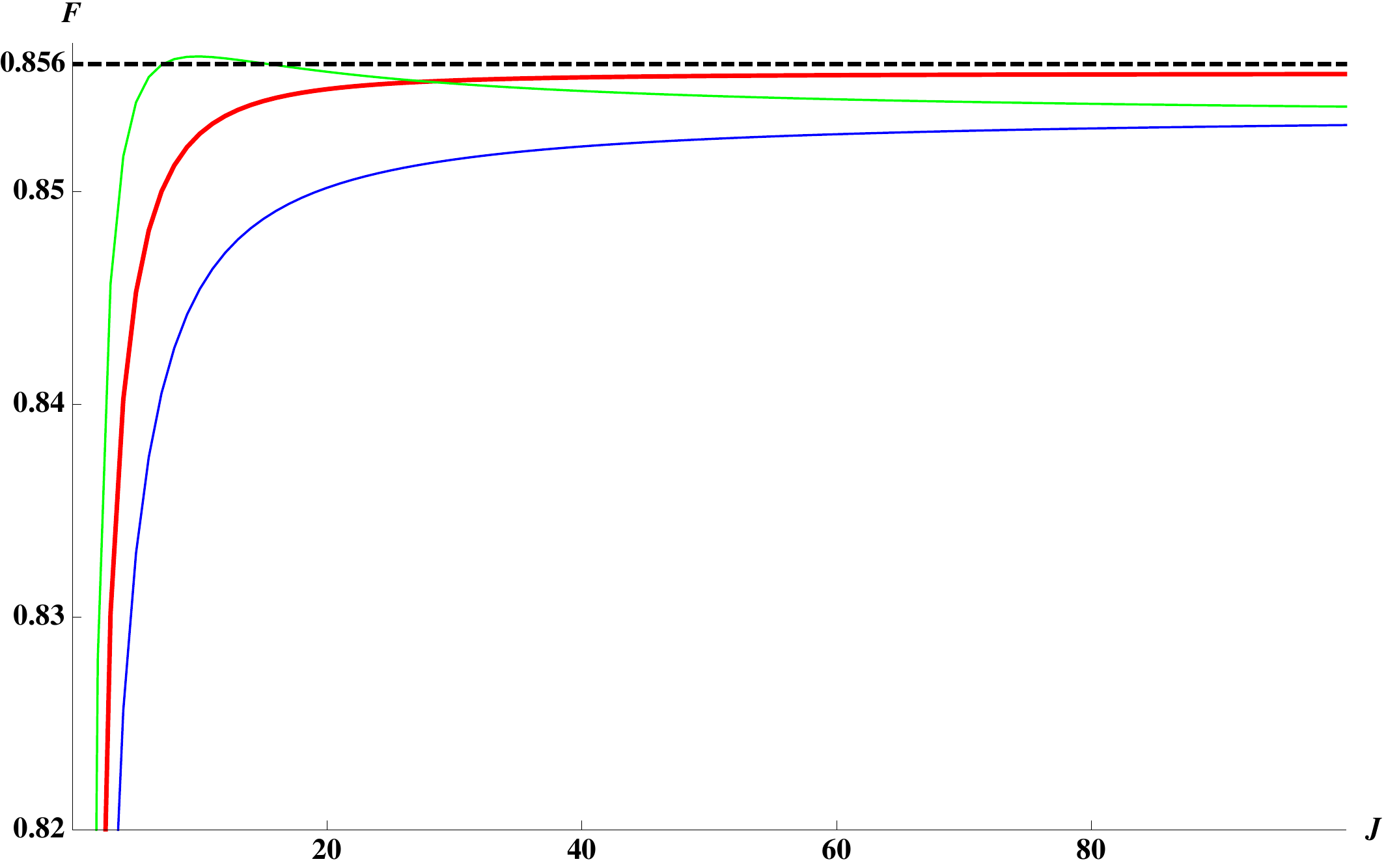}\caption{{\bf    Optimal reversible analyzer for two copies of a spin-$J$ Bell state.}   The figure shows the exact values of the fidelity in Eq. (\ref{Fdet}) as a function of $J$.  For each solid line, the number of output copies is set to $M  =  \lfloor \alpha J^2 \rfloor$, with a particular value of $\alpha$.  Specifically,  the red line represents the fidelity for the optimal value $\alpha=2.241$, which approaches $0.856$ (the dashed line) as $J$ grows large. The blue line represents the fidelity for $\alpha=2.1$, and the green line represents the fidelity for $\alpha=2.4$.}
       \label{fig:fission}
\end{figure}

To break down the two copies of the Bell state, we use the  reversible machine  defined in Proposition  (\ref{prop:isocov}).   In the large $J$ limit, this machine can produce a number of Cartesian refbits growing like $J^2$ with a non-vanish fidelity.   The exact value of the fidelity for $M =  \lfloor  \alpha  \,  J^2 \rfloor $  is plotted in Figure \ref{fig:fission}  for various values of $\alpha$.

   For large $J$,  the maximum fidelity  is obtained  for $\alpha  \approx  2.241$ and  has the value 
\begin{eqnarray}
F^{\rm iso}_{\rm Bell }   \Big[    | \Phi_{g,J}  \>^{\otimes 2}  \to   | \Phi_{g,1/2}\>^{\otimes  \lfloor 2.241 J^2 \rfloor }    \Big]    \approx 85.6\% \, , \qquad J\gg1  \, .
\end{eqnarray}   To see this,  we insert  the expressions  
\begin{eqnarray*}
\fl \qquad p^{(N,J)}_{j}=\frac{2j+1}{(2J+1)^2} \qquad {\rm and} \qquad   p^{(M,K)}_{j}=\sqrt{\frac{8(2j+1)^{4}}{\pi N^{3} }}  \exp\left[-\frac{2j^{2}}{N}\right]\left[1-O\left(\frac{j}{N}\right)\right]
\end{eqnarray*}
into the expression of the fidelity [Eq. (\ref{Fdet})].     At the leading order, we  obtain the equality
\begin{eqnarray*}
F^{\rm iso}_{\rm Bell }   \Big[    | \Phi_{g,J}  \>^{\otimes 2}  \to   | \Phi_{g,1/2}\>^{\otimes  \lfloor \alpha J^2 \rfloor }    \Big]  =  \sqrt{\frac{2\alpha^2}{\pi}}\gamma^2\left(\frac5 4,\frac 4\alpha\right),
\end{eqnarray*}
where $\gamma(s,x):=\int_0^x t^{s-1}e^{-t}\d t$ is the lower incomplete Gamma function. Maximizing  over $\alpha$, we obtain that   fidelity $85.6\%$, achieved for $\alpha=2.241$. 

\section{Probabilistic analyzer for $N$ copies of a spin-$J$ Bell state}\label{app:break2}

To obtain the desired result, we use the probabilistic machine defined by Eq. (\ref{Myes}).    The fidelity of this machine is given by Eq. (\ref{Fprob}), which in the present case becomes
\begin{eqnarray}\label{fid2}
F^{\rm prob}   \Big[    | \Phi_{g,J}  \>^{\otimes 2}  \to   | \Phi_{g,1/2}\>^{\otimes  M }    \Big] =\sum_{k=0}^{NJ}  p^{(M,1/2)}_{k} \, , \qquad N  \ge 2 \,.  
\end{eqnarray}
Now, the  probability distribution $p^{(M,1/2)}_{k}$ has the explicit form 
\begin{eqnarray*}
p^{(M,1/2)}_{k}=\frac{(2k+1)^2}{2^M(M+1)}{M+1\choose M/2+k+1} \, ,
\end{eqnarray*}
leading to the bound
\begin{eqnarray}
\nonumber F^{\rm prob}     \Big[    | \Phi_{g,J}  \>^{\otimes N}  \to   | \Phi_{g,1/2}\>^{\otimes  M }    \Big]   &= 1-\sum_{k=NJ+1}^{M/2}  p^{(M,1/2)}_{k}\\
&\ge1-(M+1)\exp\left[-\frac{2N^2J^2}{M+1}\right]  \label{boundexp}
\end{eqnarray}
following from Hoeffding's inequality \cite{hoeffding}. 

\section{A measure-and-prepare synthesizer of Bell states}\label{app:mp} 

Here we evaluate the fidelity of the measure-and-prepare synthesizer proposed in Section \ref{sec:fusion}.  

Inserting the expression of the measure-and-prepare channel  [Eqs. (\ref{MPchannel}) and (\ref{POVM})]  into the fidelity formula  (\ref{Fdef}),  we obtain the relation
\begin{eqnarray*}\label{fidmp}
\fl \qquad F^{{\rm MP}}_{\rm Bell}       \Big[    | \Phi_{g,1/2}  \>^{\otimes N}  \to   | \Phi_{g,K}\>    \Big]
&=\int \d\hat{g}~ |\<\Phi_{\hat g, K}|\Phi_{K}\>|^2~\left| \, \sum_{k=0}^{N/2}   \,        \sqrt{p^{(N,J)}_{k}}  \, \Tr\left[U_{\hat{g},k}     \right] \,  \right|^2 \, .
\end{eqnarray*}

Then, parametrizing the rotation $g$ in terms of the rotation angle (denoted by $\omega$) and the rotation axis,   we obtain  the explicit expression  
\begin{eqnarray*}
\fl \qquad F^{\rm MP}_{\rm Bell}         \Big[    | \Phi_{g,1/2}  \>^{\otimes N}  \to   | \Phi_{g,K}\>    \Big]
    =\frac{1}{\pi}\int_{-\pi}^{\pi}\d\omega~   \frac{\sin^2\left(\frac{2K+1}{2}\omega\right)}{(2K+1)^2  }~ f(\omega) \, ,
\end{eqnarray*}
where $f(\omega)$ is given by  
\begin{eqnarray*}
f(\omega):=\left|  \, \sum_{j=0}^{N/2}  ~\sqrt{p^{(N,J)}_{j}} \Tr\left[U_{\hat{g},j}\right] \, \right|^2 \, .
\end{eqnarray*}
For large $N$, the function $f(\omega)$ can be computed explicitly as
\begin{eqnarray*}
f(\omega)
&=\sqrt{\frac{\pi N^3}{2}}e^{-\frac{N\omega^2}{2}}\left[\left(\frac{\omega\cos\frac\omega2}{\sin\frac\omega2}\right)^2+\frac{2\omega\cos\frac\omega2}{N\sin\frac\omega2}   +  O(N^{-2})\right] \, ,
\end{eqnarray*}
having used   Eq. (\ref{papprox})   for the probability distribution $p_{j}^{(N,J)}$.    Note that  $f(\omega)$ decays exponentially fast for $N\omega^2\gg 1$. Hence,  we can express the measure-and-prepare fidelity as 
\begin{eqnarray}
\nonumber
\fl  \qquad  F^{\rm MP}_{\rm Bell}        \Big[    | \Phi_{g,1/2}  \>^{\otimes N}  \to   | \Phi_{g,K}\>    \Big]
& =\frac{1}{\pi}     \, \int_{-\pi  }^{\pi}   \, \d\omega~ \frac{\sin^2\left(\frac{2K+1}{2}\omega\right)}{(2K+1)^2 }~ f(\omega)  \\
\nonumber &  = \frac{\sqrt{8\pi N^3}}{\pi(2K+1)^2}   \, \int_{-\frac{1}{\sqrt{N^{1-\delta}}}}^{\frac{1}{\sqrt{N^{1-\delta}}}}\d\omega~\sin^2\left(\frac{2K+1}{2}\omega\right)e^{-\frac{N\omega^2}{2}} \,  \\
 & \quad \times  \left[1+O\left (\frac 1 N\right)\right] \, ,  \label{integration}
\end{eqnarray}
where $\delta >0$ is an arbitrary constant.  

Let us analyze the large $N$ asymptotics.   First of all, we show that the fidelity vanishes   whenever $K$ is large compared to $\sqrt N$.   This is an immediate consequence of the bound  
\begin{eqnarray*}
\fl \qquad F^{\rm MP}_{\rm Bell}          \Big[    | \Phi_{g,1/2}  \>^{\otimes N}  \to   | \Phi_{g,K}\>    \Big]
 &\le \frac{\sqrt{8\pi N^3}}{\pi(2K+1)^2}\int_{-\infty}^{+\infty}\d\omega~ e^{-\frac{N\omega^2}{2}}  \, \left[1+O\left (\frac 1 N\right)\right]  \\
&=\frac{4N}{(2K+1)^2}  \,  \, \left[1+O\left (\frac 1 N\right)\right] \, ,
\end{eqnarray*}
which follows from Eq. (\ref{integration}).  

Now, suppose that  $K$ grows as  $N^\alpha$ for some $\alpha<1/2$.   
  Picking $\delta$ so that $\alpha  < (1-\delta)/2$, we can guarantee the condition $(2K+1)\omega\ll1$  within the domain of integration in Eq. (\ref{integration}). Hence, we can  Taylor-expand $\sin^2\left(\frac{2K+1}{2}\omega\right)$, thus obtaining 
  \begin{eqnarray*}
F^{\rm MP}_{\rm Bell}          \Big[    | \Phi_{g,1/2}  \>^{\otimes N}  \to   | \Phi_{g,K}\>    \Big]
&=1-\frac{(2K+1)^2}{4N}+O(N^{-1})
\end{eqnarray*}
In conclusion, every Bell state with spin $K =  O( \sqrt N)$ can be synthesized almost perfectly in the large $N$ limit.

\section{Proof of Theorem \ref{theo:probabilisticfidelity}: analytical expression of the probabilistic fidelity}\label{app:probabilisticfidelity}

\Proof The proof uses the expression of the ultimate probabilistic fidelity derived in Ref. \cite{chiribella-xie-2013-prl}. The expression reads
\begin{eqnarray}\label{xie}
F_{\rm Bell}^{\rm prob} =\Big\|(I_{\rm out}\otimes  I_{\rm out} \otimes \tau^{-\frac12})\rho\,(I_{\rm out}\otimes I_{\rm out} \otimes\tau^{-\frac12})\Big\|_{\infty} \, ,
\end{eqnarray}
where  $\|  \cdot\|_\infty$  is the operator norm (in this case equal to the maximum eigenvalue of the operator inside the bars),    $\tau$  is the average input state 
\[\tau=\int\d g ~|\overline \Phi_{U_g}\>\<\overline \Phi_{U_g}|  \, ,\]
(the average is with respect to  the normalized Haar measure $\d g$ and the   $|\overline{\Phi}_{U_g}\>$  denotes the complex conjugate of the vector $|\Phi_{U_g}\>$),  and  
   $\rho$ is the average output-input state 
 \[\rho=\int \d g~|{\Phi}_{V_g}\>\<{\Phi}_{V_g}|\otimes|\Phi_{U_g}\>\<\Phi_{U_g}|  \, . \]

Using the double-ket notation and the isotypic decomposition  \cite{book-fulton-harris-representation} 
\begin{eqnarray*}
U_g  =  \bigoplus_{j\in{\rm Irr}  (U)}  \,   \Big ( U_g^{(j)}  \otimes I_{m_j}   \Big)    \, , 
\end{eqnarray*}
the state $|\overline\Phi_{U_g}\>$ can be expressed as 
\begin{eqnarray*}  |\overline \Phi_{U_g}  \>      &=   \frac {  | \overline U_g  \kk}{\sqrt{d_{\rm in}}}  \\
&  =    \frac  1 {\sqrt {d_{\rm in}}}  \,  \bigoplus_{j\in\set{Irr}  (U)}   \,   |\overline U_g^{(j)}  \kk  \otimes   |  I_{m_j}  \kk  \, ,           
\end{eqnarray*}
where we reordered the tensor factors in order to have the two representation spaces on the left and the two multiplicity spaces on the right.  
Applying Schur's lemma \cite{book-fulton-harris-representation}, the average state $\tau$  be explicitly calculated as 
\begin{eqnarray*}
\tau=\frac 1 {d_{\rm in}}  \, \bigoplus_{j\in\set{Irr} (U)} \frac{I_{j}\otimes I_{j}\otimes |I_{m_j}\kk\bb I_{m_j}|}{d_j} \, ,
\end{eqnarray*}
where $d_{\rm in}$ is the dimension of the input Hilbert space and $I_j$ is the identity matrix on the representation space where the representation  $\{U_g^{(j)}\}$ acts.   The inverse square root $\tau^{-1/2}$ is then given by  
\begin{eqnarray*}
\tau^{-1/2}=\sqrt{d_{\rm in}  }  \, \bigoplus_{j\in\set{Irr} (U)}    \sqrt{\frac{d_j}{m_j}  }\frac{I_{j}\otimes I_{j}\otimes |I_{m_j}\kk\bb I_{m_j}|}{m_j} \, .
\end{eqnarray*}
Hence, we have the relation   
\begin{eqnarray}
\nonumber
&    I_{\rm out}  \otimes  I_{\rm out}  \otimes  \tau^{-\frac 12} =  \\
\nonumber 
&  =   \sqrt{d_{\rm in}  }  \, \bigoplus_{j\in\set{Irr} (U)}    \sqrt{\frac{d_j}{m_j}  }\frac{  I_{\rm out} \otimes I_{\rm out} \otimes I_{j}\otimes I_{j}\otimes |I_{m_j}\kk\bb I_{m_j}|}{m_j}   \\
 &  =   \sqrt{d_{\rm in}  }  \, \bigoplus_{j\in\set{Irr} (U)}   \bigoplus_{l,l'\in  \set{Irr}  \left(  V \otimes \overline U^{(j)}\right)} \sqrt{\frac{d_j}{m_j}  }\frac{  I_{l} \otimes I_{l'} \otimes I_{m_l^{(j)}}  \otimes I_{m_{l'}^{(j)}}  \otimes |I_{m_j}\kk\bb I_{m_j}|}{m_j} \qquad  \qquad     
  \label{I*tau} 
\end{eqnarray}
where again we reordered the tensor factors in order to have all the representation spaces on the left and all the multiplicity spaces on the right.

Computing the average state $\rho$ is a bit more complex.    First, we express the product state  $|{\Phi}_{V_g}\>  \otimes|\overline\Phi_{U_g}\>$  as  
\begin{eqnarray}
\nonumber |{\Phi}_{V_g}\>  \otimes|\overline\Phi_{U_g}\> &  =  \frac {  |  V_g\kk}{\sqrt {d_{\rm out}}} \otimes  \frac {  |  \overline U_g\kk}{\sqrt {d_{\rm in}}}   \\
\nonumber &  =   \frac 1  {\sqrt{d_{\rm in }  d_{\rm out}  }}  \bigoplus_{j\in\set{Irr} (U) }   \,     |  V_g\kk   \otimes   |\overline   U^{(j)}_g\kk  \otimes  |I_{m_j}  \kk      \\
&  =   \frac 1  {\sqrt{d_{\rm in}   d_{\rm out}}}  \bigoplus_{j\in\set{Irr} (U) }  \bigoplus_{l\in\set{Irr} \left(V\otimes \overline U^{(j)}\right) }     \,   |   U^{(l)}_g\kk       \otimes  | I_{m_l^{(j)}}  \kk \otimes  |I_{m_j}  \kk  \, ,  \qquad \qquad \label{equazione}
\end{eqnarray}
with the usual reordering of the tensor factors.   Here, $m_l^{(j)}$ is the multiplicity of the representation $\{  U_g^{(l)}\}$ in the decomposition of the representation $\{  V_g\otimes \overline  U_g^{(j)} \}$.    

Note that we  have  $m_l^{(j)}  =  0$ when 
 the representation  $\{  U_g^{(l)}\}$ does not appear in the decomposition of   $\{  V_g\otimes \overline  U_g^{(j)} \}$.
  In this case, we adopt the convention $I_{m_l^{(j)}}  =  0$, which allows us to  exchange the order of the two summations in Eq. (\ref{equazione}), obtaining 
\begin{eqnarray*}
|{\Phi}_{V_g}\>  \otimes|\overline\Phi_{U_g}\> 
&  =   \frac 1  {\sqrt{d_{\rm in}   d_{\rm out}}}   \bigoplus_{l\in\set{Irr} (V\otimes \overline U) }   \bigoplus_{j\in\set{Irr} (U) }     \,   |   U^{(l)}_g\kk       \otimes  | I_{m_l^{(j)}}  \kk \otimes  |I_{m_j}  \kk  \\
&   =   \frac 1  {\sqrt{d_{\rm in}   d_{\rm out}}}  \bigoplus_{l\in \set{Irr}  \left( {V}    \otimes \overline U  \right) }       \,    |  U^{(l)}_g\kk   \otimes      |\alpha_l\> \, ,
\end{eqnarray*}
with  
\begin{eqnarray}\label{alphal} 
|\alpha_l\>   :  =    \bigoplus_{j\in\set{Irr} (U) }   \,        |I_{m_{l}^{(j)}} \kk    \otimes  |I_{m_j}  \kk   \, ,
\end{eqnarray}

 We are now ready to compute the average state $\rho$, which reads
\begin{eqnarray}
\nonumber  \rho  &=\int \d g~|\overline{\Phi}_{V_g}\>\<\overline{\Phi}_{V_g}|\otimes|\Phi_{U_g}\>\<\Phi_{U_g}|\\
\label{rho'}
  &  =   \frac 1  {{d_{\rm in}   d_{\rm out}}}     \bigoplus_{l\in   \set {Irr}\left( V\otimes \overline U  \right) }       \,     \frac{  I_l  \otimes  I_l  \otimes  |\alpha_l   \>\<\alpha_l |  }{d_l}   \, . 
\end{eqnarray}

Combining  with Eqs.   (\ref{I*tau}) and  (\ref{rho'}), we obtain the relation
\begin{eqnarray}
\nonumber  &(I_{\rm out}\otimes  I_{\rm out} \otimes \tau^{-\frac12})\rho\,(I_{\rm out}\otimes I_{\rm out} \otimes\tau^{-\frac12})  \\
  \label{bastissima} &  =  \frac 1  {{  d_{\rm out}}}     \bigoplus_{l\in   \set {Irr}\left( V\otimes \overline U  \right) }        \,    \frac{  I_l  \otimes  I_l  \otimes   |\alpha'_l\>\<\alpha'_l| }{d_l} \, ,
\end{eqnarray}
with 
\begin{eqnarray*}
|\alpha_l'\>  :  =     \bigoplus_{j\in\set{Irr} (U) }    \,  \sqrt{\frac{ d_j}{m_j}}     | I_{m_{l}^{(j)}} \kk  \otimes  |I_{m_j}  \kk   \, .  
\end{eqnarray*}
We now reached the conclusion.  In order to compute the fidelity, Eq. (\ref{xie}) tells us that we must compute the maximum eigenvalue of the operator in Eq. (\ref{bastissima}).    The eigenvalues are  
\begin{eqnarray*}  \lambda_{l}    &  =      \frac {\< \alpha_l'|  \alpha_l'\>}{d_{\rm out} d_l}  \\
&  =  \sum_{j\in\set{Irr} (U) }  \frac{ d_j  \, m_l^{(j)}}{d_{\rm out}  d_l} \, .
\end{eqnarray*}
Maximizing over $l$ one then obtains  the desired expression  
      \begin{equation}
      \label{mortina}
      F_{\rm Bell}^{\rm prob}\left(   |\Phi_{U_g} \> \to    |\Phi_{V_g} \> \right)  = 
       \max_{l \in  {\rm Irr}  ( V\otimes \overline U)}   \left[  \frac 1{  d_{\rm out}  d_l}   \left(    \sum_{j\in\set{Irr} (U) }   \,    d_j  \,m_l^{(j)}   \right)  \right]  \, .   
 \end{equation}   
   \qed

\section{Proof of Theorem \ref{theo:prob=det}: irreducibility implies no probabilistic advantage}\label{app:prob=det}

\Proof  Let us start from the case of the Bell state conversion.   We use  a general result from Ref. \cite{chiribella-yang-2013-natcomm}, stating that the optimal probabilistic and deterministic operations perform equally well whenever the set of input states is invariant under the action of an irreducible group representation.    Thanks to this result, we only need to show that the set of input states  $\{  |\Phi_{U_g}  \>  \}$ is invariant under the action of an irreducible representation.  In our case, the irreducible representation is   $ \{   U_h \otimes \overline U_k\}$, where the elements $h$ and $k$ vary independently over the group $\grp G$, and $\overline U_k$ is the complex conjugate of $U_k$ with respect to a fixed basis, regarded as the ``computational basis".   The irreducibility of the representation  $ \{   U_h \otimes \overline U_k\}$ is immediate from the assumption that the representation $\{  U_g\}$ is irreducible.  

The invariance of the set $\{  |\Phi_{U_g}  \>  \}$ is immediate from the relation  
\[   (U_h \otimes \overline U_k)\,   |\Phi_{U_g}\>   =    \frac{   |  U_h U_g  U_k^\dag\,  \kk}{\sqrt{d_{\rm in}}}     =    |\Phi_{ U_{hgk^{-1}}}  \> \, ,  \]
valid for arbitrary group elements $g,k,$ and $h$.   Hence, the result of Ref. \cite{chiribella-yang-2013-natcomm} guarantees that there is no difference in performance between probabilistic and deterministic operations.

\begin{figure}[h!]
\centering
      \includegraphics[width=0.5\textwidth]{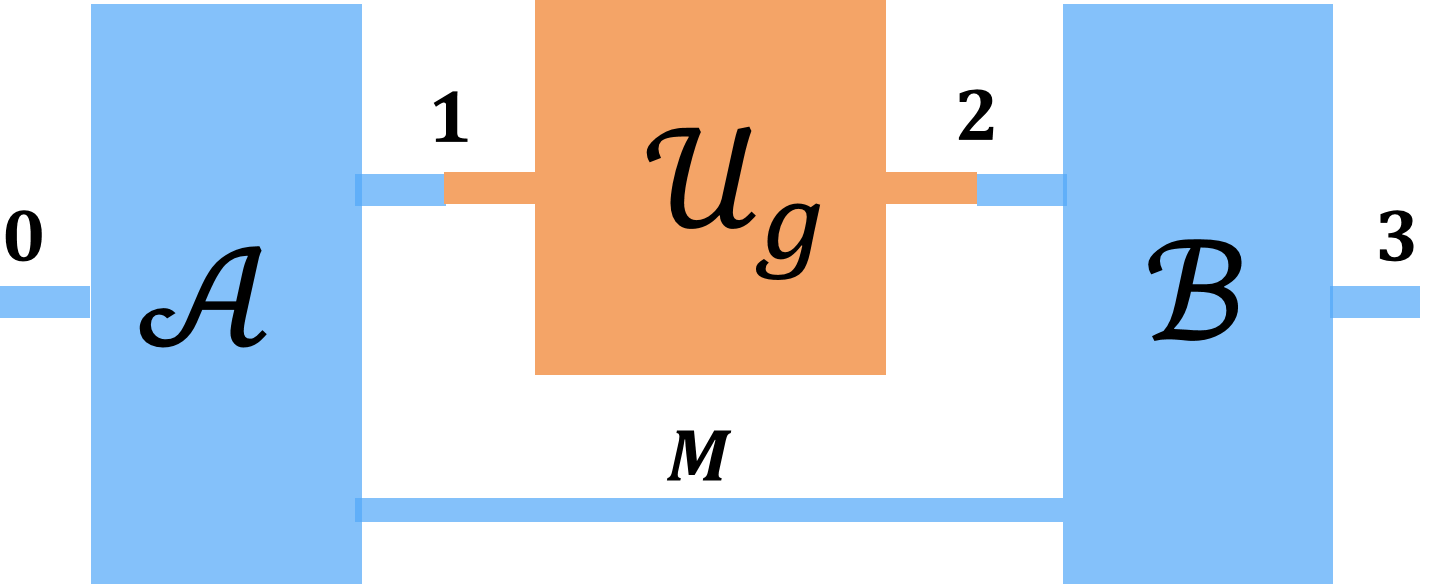}\caption{{\bf   Gate simulation network.}   The figure illustrates the general form of a network connected with a black box implementing the  gate $U_g$. Here $\map{A}$ and $\map B$ are generic quantum operations, heralded by the outcomes of two probabilistic processes. }
       \label{fig:circuit}
\end{figure}

Let us consider now the case of gate simulations, implemented by quantum networks of the form of Figure \ref{fig:circuit}.    A convenient way to describe the gate simulation network is to use the method of quantum combs \cite{chiribella2008quantum,chiribella2009theoretical}, which associates the  network with a positive operator $R$  acting on  the four Hilbert spaces 
\begin{eqnarray*}
\spc H_0   \simeq   \spc H_{\rm out} \, , \qquad   \spc H_1  \simeq  \spc H_{\rm in}  \, , \qquad  \spc H_2   \simeq  \spc H_{\rm in}  \, ,\qquad \spc H_3  \simeq  \spc H_{\rm out} \, ,     
\end{eqnarray*}
where $\spc H_{\rm in}$ is the space on which the gate $U_g$ acts, and $\spc H_{\rm out}$ is the gate on which $V_g$ acts.  
In terms of the operator $R$,    the probabilistic fidelity  is given by  
\begin{eqnarray}\label{combazio}
F_{\rm gate}^{\rm prob}   =  \frac { \int \d g \,       \Big (\bb  V_g|_{30}  \otimes \bb \overline U_g |_{21}\Big)  \,  R  \,\Big (   |V_g\kk_{30}\otimes  |\overline U_g  \kk_{21} \Big)}{ \int \d g \,   \Tr_{30}  \Big[  \bb   \overline U_g |_{21}  \,   R \,  |  \overline  U_g\kk_{21}\Big]}  \, ,
\end{eqnarray}      
where the subscripts identify the Hilbert spaces to which the vectors belong, and $\Tr_{30}$ denotes the partial trace over the Hilbert space $\spc H_3\otimes \spc H_0$. 
 The symmetry of Eq. (\ref{combazio}) implies that, without loss of generality,  the optimal network can be chosen with an operator $R$ satisfying the condition 
  \begin{eqnarray}\label{combsymm}
 [R,    V_{h,3}  \otimes  \overline U_{h,2}  \otimes U_{k, 1}    \otimes  \overline V_{k,0}   ]  =  0  \, , \qquad \forall  h \in  \grp G \, , \forall  k\in\grp G \, .
 \end{eqnarray}
 Now,   Eq. (\ref{combsymm}) implies that the network has the same performance of a deterministic network.   To prove this, we recall that a network is deterministic if and only if its operator $R$ satisfies the conditions  \cite{chiribella2008quantum,chiribella2009theoretical}
 \begin{eqnarray}\label{combdeterministico}
 \Tr_3  [  R  ]  =     I_2\otimes  R'   \qquad {\rm and}  \qquad \Tr_1  [  R']   =    I_0  \, ,      
 \end{eqnarray} 
 for some nonnegative operator $R'$ acting on $\spc H_1\otimes \spc H_0$.    Now, Eq. (\ref{combsymm}) yields the condition  
 \[  [\Tr_3 [R]   ,  \overline U_{h,2} \otimes U_{k, 1}    \otimes  \overline V_{k,0}       ]  =  0  \, , \qquad \forall  h \in  \grp G \, , \forall  k\in\grp G \, ,   \] 
 which in turn implies the condition  
 \[  \Tr_3 [ R]  =     I_2\otimes  R'    \]
 having used Schur's lemma and the fact that $\{  U_h\}$ is an irreducible representation.  Moreover, the operator $R'$ must satisfy the condition   
   \begin{eqnarray*}
  [ R',     U_{k, 1}  \otimes \overline V_{k,0}    ]  =  0  \, , \qquad  \forall  k\in\grp G \, ,    
 \end{eqnarray*}
 which implies  
   \[ [ \Tr_1[R'],     \overline V_{k, 0}     ]  =  0  \, , \qquad  \forall  k\in\grp G \, , \] 
  and, in turn,  
  \[  \Tr_1 [R']   =  \lambda\,  I_0 \, ,\]
  for some constant $\lambda$.    The last equality is a consequence of Schur's lemma, applied to the irreducible representation $\{  \overline V_k\}$.      
  
  Defining $R^{\rm det}  :  =  R/\lambda$,  we then have that $R_{\rm det}$ is the operator of a deterministic quantum network.  
  From Eq. (\ref{combazio}) it is immediate that the deterministic network with operator  $R_{\rm det}$ has the same fidelity of the probabilistic network with operator $R$.  \qed  
   
\section{Proof of Theorem \ref{theo:local}:  when local operations and memoryless networks are optimal}  

\Proof   Let us start from the case of the Bell state conversion.  The proof uses  the Choi isomorphism, which associates  the quantum channel $\map C$ to the  operator   $C$, acting on $\spc H_{\rm out1}  \otimes  \spc H_{\rm out2}  \otimes  \spc H_{\rm in1}  \otimes \spc H_{\rm in2}  $,   where the subscripts 1 and 2 label the two spaces in the input and output Bell pairs. 
 Note that, in order to be the Choi operator of a quantum channel, the operator $C$ must be positive and must satisfy the normalization condition 
\begin{equation}\label{choinorm}\Tr_{\rm out 1  ,  out 2} [C]  = I_{\rm in1}  \otimes  I_{\rm in2}  \, . 
\end{equation}
In terms of the Choi operator, the fidelity for the Bell state conversion $|\Phi_{U_g}\>  \to |\Phi_{V_g}\>$ can be written as \cite{chiribella-xie-2013-prl}   
\[F_{\rm Bell}   =  \Tr  [   C   \, \rho]   \, ,\] where $\rho$ is the   state  
\begin{eqnarray}
\nonumber  \rho  &=\int \d g~|\overline{\Phi}_{V_g}\>\<\overline{\Phi}_{V_g}|\otimes|\Phi_{U_g}\>\<\Phi_{U_g}|\\
\label{rhosec}
  &  =   \frac 1  {{d_{\rm in}   d_{\rm out}}}     \bigoplus_{l\in   \set {Irr}\left( V\otimes \overline U  \right) }       \,     \frac{  I_l  \otimes  I_l  \otimes  |\alpha_l   \>\<\alpha_l |  }{d_l}   \, ,   \qquad {\rm with}  \qquad   \<\alpha_l | \alpha_l\>  =  m_l  \qquad   \qquad
\end{eqnarray}
cf. Eq. (\ref{rho'}).     Recall that there is no difference between the probabilistic fidelity and the deterministic fidelity, since the representation $\{ U_g\}$ is irreducible  (cf.  Theorem \ref{theo:prob=det}).  Hence,   the maximum fidelity    for the deterministic Bell state conversion $|\Phi_{U_g}\>  \to |\Phi_{V_g}\>$  is given  by Eq. (\ref{mortenera}),    which provides the expression 
\begin{eqnarray}
F^{\rm det}_{\rm Bell}     =   \frac{d_{\rm in}}{d_{\rm out}} \,    \left[  \max_{l \in  {\rm Irr}  (  V \otimes \overline U)}  \,    \frac{m_l}{ d_l}   \right]  \, ,  
\end{eqnarray}
where $m_l$ is the multiplicity of the representation $\{  U_g^{(l)}\}$ in the decomposition of $\{  V_g \otimes \overline U_g\}$.  By direct inspection, one can see that the optimal fidelity is attained by  the operator 
\[  C_{\rm opt}   =    d^2_{\rm in}   \,   \left ( \,    \frac {I_{l_*}}{d_{l_*}}   \otimes \frac{I_{l_*}}{d_{l_*}}   \otimes   \frac{|\alpha_{l_*}  \>\< \alpha_{l_*} |}{\< \alpha_{l_*} |\alpha_{l_*}  \>} \,  \right) \, ,  \]
where $l_*$ is the value of $l$ that maximizes the ratio $m_l/d_l$.  Note that the operator $C_{\rm opt}$
 is positive and satisfies Eq. (\ref{choinorm}):  indeed, one has  
 \begin{eqnarray}
\nonumber  
\Tr_{\rm out1, out2  }  [C]  &=  d_{\rm in}^2 \,        \Tr_{\rm out 2}  \left[   \Tr_{\rm out 1}  \left[      \frac {I_{l_*}}{d_{l_*}}   \otimes \frac{I_{l_*}}{d_{l_*}}       \otimes   \frac{|\alpha_{l_*}  \>\< \alpha_{l_*} |}{\< \alpha_{l_*} |\alpha_{l_*}  \>} \right] \right]  \\
&=  d_{\rm in} \,            I_{\rm in1}  \otimes  \,  \Tr_{\rm out 2}  \left[  \frac{I_{l_*}}{d_{l_*}}       \otimes   \Tr_{\spc M_1}  \left[  \frac{|\alpha_{l_*}  \>\< \alpha_{l_*} |}{\< \alpha_{l_*} |\alpha_{l_*}  \>} \right] \right]  \, ,
 \end{eqnarray} 
 where we applied Schur's lemma to the irreducible representation $\{ U_g\}$, and $\Tr_{\spc M_1}$ denotes the partial trace over the multiplicity spaces resulting from the coupling of first  systems in the input and output Bell pairs.
 Finally, we apply again Schur's lemma to the irreducible representation $\{ U_g\}$, obtaining 
  \begin{eqnarray}
\nonumber  
\Tr_{\rm out1, out2  }  [C]  
&=  d_{\rm in} \,            I_{\rm in1}  \otimes  \,  \Tr_{\rm out 2}  \left[  \frac{I_{l_*}}{d_{l_*}}       \otimes   \Tr_{\spc M_1}  \left[  \frac{|\alpha_{l_*}  \>\< \alpha_{l_*} |}{\< \alpha_{l_*} |\alpha_{l_*}  \>} \right] \right] \\
\nonumber  
 &=       I_{\rm in1}  \otimes         I_{\rm in2}  \,    \Tr_{\spc M_1 \otimes \spc M_2}  \left[  \frac{|\alpha_{l_*}  \>\< \alpha_{l_*} |}{\< \alpha_{l_*} |\alpha_{l_*}  \>}  \right] \\
\nonumber  
 &=       I_{\rm in1}  \otimes         I_{\rm in2}  \,    \Tr  \left[  \frac{|\alpha_{l_*}  \>\< \alpha_{l_*} |}{\< \alpha_{l_*} |\alpha_{l_*}  \>} \right]  \\
 &=       I_{\rm in1}  \otimes         I_{\rm in2}  \, .
 \end{eqnarray} 
   Hence, $C_{\rm opt}$ is the Choi operator of a quantum channel.   If $m_{l_*}=1$,  the Choi operator $C_{\rm opt}$  has the product form  $C_{\rm opt}   =    A \otimes B $,  with
    \[  A  =   B  =   \frac {d_{\rm in}}{d_{l_*}}   \,  I_{l_*} \,.  \]  
    Here,  $A$ is the Choi operator of  a channel $\map A$ transforming system $\rm in 1$ into system $\rm out 1$, while  $B$ is the Choi operator of a channel $\map B$ transforming system $\rm in2$ into system $\rm out 2$.    In conclusion, the optimal Bell state conversion is implemented with local operations performed independently on the two systems of the input Bell pair.   
 
 Let us consider now the case of the gate simulation.    The gate simulation network is described by a quantum comb $R$, which can be chosen without loss of generality to satisfy the commutation relation  (\ref{combsymm}).  In terms of the quantum comb $R$,  the fidelity can be written as  
 \[ F_{\rm gate}  =  \frac {d_{\rm in}}{d_{\rm out}} \, \Tr  [  R   \rho]  \,  , \]
 where $\rho $ is the state  in Eq. (\ref{rhosec}).  Again, there is no difference between probabilistic and deterministic fidelity, because the representations $\{  U_g\}$ and  $\{V_g\}$ are both irreducible  (cf. Theorem \ref{theo:prob=det}). Hence, the maximum fidelity is provided by Eq. (\ref{mortenera}), which yields
  \begin{eqnarray}
F^{\rm det}_{\rm gate}     =   \frac{d_{\rm in}}{d_{\rm out}} \,    \left[  \max_{l \in  {\rm Irr}  (  V \otimes \overline U)}  \,    \frac{m_l}{ d_l}   \right]  \, ,  
\end{eqnarray}
  By direct inspection, we  find that the optimal fidelity is attained by the operator  
 \[  R_{\rm opt}   =    d_{\rm in}  d_{\rm out}     \,   \left ( \,    \frac {I_{l_*}}{d_{l_*}}   \otimes \frac{I_{l_*}}{d_{l_*}}   \otimes   \frac{|\alpha_{l_*}  \>\< \alpha_{l_*} |}{\< \alpha_{l_*} |\alpha_{l_*}  \>} \,  \right) \, ,  \]
where $l_*$ is the value of $l$ that maximizes the ratio $m_l/d_l$.  The operator $R$ represents a deterministic quantum network, because the conditions  (\ref{combdeterministico}) are satisfied:  
Indeed, one has 
\begin{eqnarray}
 \nonumber \Tr_3 [ R]     & =   d_{\rm in}  d_{\rm out}     \,   \Tr_3    \left [ \,    \frac {I_{l_*}}{d_{l_*}}   \otimes \frac{I_{l_*}}{d_{l_*}}   \otimes   \frac{|\alpha_{l_*}  \>\< \alpha_{l_*} |}{\< \alpha_{l_*} |\alpha_{l_*}  \>} \,  \right] \\  
\label{ccab} & =     d_{\rm out}     \,   I_2  \otimes   \frac{I_{l_*}}{d_{l_*}}   \otimes   \Tr_{\spc M_{32}}  \left[\frac{|\alpha_{l_*}  \>\< \alpha_{l_*} |}{\< \alpha_{l_*} |\alpha_{l_*}  \>} \,  \right]    \, ,  
\end{eqnarray}
   where we applied Schur's lemma to the irreducible representation $\{  U_g\}$ and we used the notation $\Tr_{\spc M_{32}}$ to denote the partial trace over the multiplicity spaces resulting from the coupling of systems $3$ and $2$.    
  Eq. (\ref{ccab}) implies that we have $\Tr_3 [R]  =   I_{2}\otimes R'$ with  
   \[  R'  =   d_{\rm out}  \,  \frac{I_{l_*}}{d_{l_*}}   \otimes   \Tr_{\spc M_{32}}  \left[\frac{|\alpha_{l_*}  \>\< \alpha_{l_*} |}{\< \alpha_{l_*} |\alpha_{l_*}  \>} \,  \right] \, . \]
   Moreover, we have 
   \begin{eqnarray}
  \nonumber  \Tr_1 [  R']   &  =   d_{\rm out}  \,  \Tr_1  \left[   \frac{I_{l_*}}{d_{l_*}}   \otimes   \Tr_{\spc M_{32}}  \left[\frac{|\alpha_{l_*}  \>\< \alpha_{l_*} |}{\< \alpha_{l_*} |\alpha_{l_*}  \>} \,  \right]  \right]  \\
   \nonumber &   =   I_{0}  \,      \Tr_{\spc M_{32}\otimes \spc M_{10}}  \left[\frac{|\alpha_{l_*}  \>\< \alpha_{l_*} |}{\< \alpha_{l_*} |\alpha_{l_*}  \>} \,  \right]    \\
 \nonumber   &  =  I_{0}  \,      \Tr  \left[\frac{|\alpha_{l_*}  \>\< \alpha_{l_*} |}{\< \alpha_{l_*} |\alpha_{l_*}  \>} \,  \right]  \\
   &  =  I_0  \, ,
   \end{eqnarray}
   the second equality following from Schur's lemma applied to the representation $\{ V_g\}$.  This concludes the proof that $R$ represents a deterministic quantum network.  
   
  If $m_{l_*}=1$,  the  quantum comb  $R_{\rm opt}$  has the product form  $R_{\rm opt}   =    d_{\rm in} d_{\rm out}  \,   \left ( \,    \frac {I_{l_*}}{d_{l_*}}   \otimes \frac{I_{l_*}}{d_{l_*}}\right) $, where the first factor acts on the Hilbert spaces $\spc H_3$ and $\spc H_2$, while the second acts on the Hilbert spaces $\spc H_1$ and $\spc H_0$.   This  means  that the optimal network consists of  a quantum channel $\map A$ from system $0$ to system $1$, followed by a quantum channel  $\map B$ from system $2$ to system $3$, the two channels having the Choi operators 
  \[  A  =     \frac   {d_{\rm in}}{d_{l_*}}{  I_{l_*}}  \qquad {\rm and}  \qquad  B  =  \frac  {  d_{\rm out}}{d_{l*}} \,  I_{l_*} \, ,\]
  respectively.   Note that no quantum memory is needed between $\map A$ and $\map B$.   \qed

\section{Proof of Theorem \ref{theo:nontrivial}: lower bound on the gate fidelity}\label{app:gate}  

\Proof We have to prove the  bound 
$$ F_{\rm gate}^{\rm det}  \Big [  U_g\to V_g  \Big]\ge \left(F_{\rm Bell}^{\rm prob}  \Big [   |\Phi_{U_g}  \>  \to  |\Phi_{V_g}  \>    \Big]\right)^2 \, , $$
where  $F_{\rm gate}^{\rm det}$ is the ultimate deterministic fidelity of the gate simulation $U_g\to V_g$ and  $F_{\rm Bell}^{\rm prob}$ is the optimal fidelity of the corresponding spin conversion $|\Phi_{U_g}\>\to |\Phi_{V_g}\>$. 

To derive the bound we start from the decomposition of the representations $\{  U_g\}$ and $\{V_g\}$. Explicitly, we write  
\begin{eqnarray}\label{decuno}
U_g=\bigoplus_{j\in\set{Irr}  (U)} \left(U_g^{(j)}\otimes I_{m_j}\right)\qquad   {V}_g=\bigoplus_{k\in\set{Irr} ( V)} \left({U}_g^{(k)}\otimes I_{n_k}\right) \, .
\end{eqnarray}
In addition, we decompose the representation   $U_g^{(k)}\otimes\overline{U}_g^{(j)}$    as 
\begin{eqnarray}\label{decdue}
U_g^{(k)} \otimes\overline{U}_g^{(j)}=\bigoplus_{l\in\set{Irr} \left (   U^{(k)}\otimes \overline U^{(j)}\right)} 
\left(U_g^{(l)}\otimes I_{m_{l}^{(k,j)}}\right)\label{character} \, ,
\end{eqnarray}
where $m_l^{(k,j)}$ is the multiplicity of the irreducible representation $\{  U^{(l)}_g\}$ in the decomposition of the product representation $\{  U_g^{(k)} \otimes \overline{U}_g^{(j)} \}$.     The dimensions of the representation and multiplicity spaces in Eqs.  (\ref{decuno}) and (\ref{decdue})  determine the optimal fidelity.  The exact formula is given by   Proposition 5 of \cite{bisio-dariano-2014-pla},  which yields the expression
\begin{eqnarray*}
F_{\rm gate}^{\rm det}=\max_{\{h_{k,l}\}}\sum_{l\in  \set{Irr}  (  V\otimes \overline U)}\left(\sum_{k\in\set{Irr}  (V)}\sqrt{\sum_{j\in\set{Irr}  (U)}\frac{n^2_k d_k d_{j}m_j m_l^{(k,j)}}{d_l d_{\rm out}^2}\cdot h_{k,l}}\right)^2,
\end{eqnarray*}
where $\{h_{k,l}\}$ is a set of coefficients satisfying the constraints $h_{k,l}\ge0$ and $\sum_{l}h_{k,l}=1\,\forall k$.
Choosing $h_{k,l}=\delta_{ll_*}$ for some fixed (but otherwise arbitrary) $l_*$, we get a lower bound of the gate simulation fidelity. Specifically, we have
\begin{eqnarray}
F_{\rm gate}^{\rm det}  \Big [  U_g\to V_g  \Big]  &\ge \max_{l_* \in \set {  Irr}  (  V\otimes \overline U)}\left(\sum_{k\in\set{Irr} (V)}q_k\sqrt{\sum_{j\in\set{Irr}  (U)} \frac{d_j  m_jm^{(k,j)}_{l_*}}{d_kd_{l_*}}}\right)^2\qquad q_k:=\frac{d_k n_k}{d_{\rm out}}\nonumber\\
&\ge
 \max_{l_*  \in  \set{Irr}  (   V\otimes \overline U)}\left(\sum_{k\in\set{Irr}  (V)}q_k\sqrt{\sum_{j\in\set{Irr} (U)}\frac{ d_j  m^{(k,j)}_{l_*}}{d_kd_{l_*}}}\right)^2 \, .\label{app-final1}
\end{eqnarray}
Now, we use the standard group-theoretic formula for the multiplicities  \cite{book-fulton-harris-representation}, which can be computed as 
\begin{eqnarray}\label{character1}
m^{(k,j)}_{l}   =  \int \d g\,  \overline{\chi}_g^{(l)}    \chi_g^{(k)}   \overline \chi_g^{(j)}
\end{eqnarray}
in terms of the characters   $\chi_{g}^{(j)}  :=  \Tr  [ U_g^{(j)}]$, $\chi_g^{(k)} :=  \Tr[ U_g^{(k)}]$ and $\chi_g^{(l)}: =  \Tr [U_g^{(l)}] $. 

A trivial rearrangement of the terms gives
\begin{eqnarray}
\nonumber m^{(k,j)}_{l}     &=  \int \d g\,  \overline{\chi}_g^{(l)}    \chi_g^{(k)}   \overline \chi_g^{(j)}  \\ 
  \nonumber & =    \int \d g\,    \overline \chi_g^{(j)}  
        \chi_g^{(k)}   \overline{\chi}_g^{(l)}  \\
             &  =  m_j^{(k,l)} \, ,
        \end{eqnarray}
 where $  m_j^{(k,l)}$ is the multiplicity of the irreducible representation $\{  U_g^{(j)}\}$ in the decomposition of the product representation $\{  U_g^{(k)} \otimes \overline  U_g^{(l)}\}$. 
 
 On the other hand, the decomposition  
 \begin{equation}
 \spc R_k \otimes \spc R_l  =   \bigoplus_{j \in  \set{Irr}  \left(   U^{(k)}  \otimes  \overline U^{(l)}\right)} \,    \left(  \spc R_j \otimes \spc M_j^{(k,l)} \right) 
 \end{equation}
 implies the relation 
 \begin{eqnarray*}
d_k  d_l =\sum_{j \in \set{Irr}  \left(   U^{(k)}  \otimes  \overline U^{(l)}\right)  } d_j m_j^{(k,l)}=\sum_{j \in  \set{Irr}  \left(   U^{(k)}  \otimes  \overline U^{(l)}\right)} d_j m_l^{(k,j)},
\end{eqnarray*}
which in turn implies the inequality
\begin{eqnarray*}
\frac{\sum_{j \in\set{Irr}  (U)}d_j  m_l^{(k,j)}}{d_kd_l}  \le  \frac{\sum_{j \in\set{Irr}  \left(U^{(k)}\otimes \overline U^{(l)}\right)}d_j  m_l^{(k,j)}}{d_kd_l}  =1 \, ,
\end{eqnarray*}
and therefore  
\begin{eqnarray*}
\sqrt {\frac{\sum_{j \in\set{Irr}  (U)}d_j  m_l^{(k,j)}}{d_kd_l}}  \ge  \frac{\sum_{j \in\set{Irr}  (U)}d_j  m_l^{(k,j)}}{d_kd_l}   \, .
\end{eqnarray*}
Using the above property, we can reduce the fidelity bound (\ref{app-final1}) to
\begin{eqnarray*}
F_{\rm gate}^{\rm det}  \Big [  U_g\to V_g  \Big]    &  \ge     \max_{l_* \in  \set{Irr}  (   V\otimes \overline U)}\left(\sum_{k\in\set{Irr}  (V)}q_k\sqrt{\sum_{j\in\set{Irr} (U)}\frac{ d_j  m^{(k,j)}_{l_*}}{d_kd_{l_*}}}\right)^2  \\
  &\ge \max_{l_*  \in  \set{Irr}  (   V\otimes \overline U) }\left(\sum_{k\in\set{Irr}  (V)}q_k\sum_{j\in\set{Irr}  (U)}\frac{d_j m_{l_*}^{(k,j)}}{d_kd_{l_*}}\right)^2\\
  &=   \max_{l_* \in  \set{Irr}  (   V\otimes \overline U)}\left(  \frac 1 {d_{\rm out}}\,    \sum_{j\in\set{Irr}  (U)}  \,  d_j  \sum_{k\in\set{Irr}  (V)}  \,  \frac{m_{l_*}^{(k,j)} m_k  }{d_{l_*}}\right)^2\\
   &=   \max_{l_* \in  \set{Irr}  (   V\otimes \overline U)}\left[  \frac 1 {d_{\rm out}  d_{l_*}}\,  \left(    \sum_{j\in\set{Irr}  (U)}  d_j  m_{l_*}^{(j)} \right)\right]^2 \, , 
\end{eqnarray*}
where $m_{l_*}^{(j)}$ is the multiplicity of the irreducible representation $\{  U_g^{(l_*)}\}$ in the decomposition of the product representation $  \{  V_g\otimes  \overline U_g^{(j)}  \}$.   
The last term in the inequality is exactly the fidelity of the Bell state conversion, as given by Eq.   (\ref{mortina}). \qed

\end{document}